\documentclass[%
 reprint,
superscriptaddress,
 amsmath,amssymb,
 aps,
prb,
]{revtex4-1}

\usepackage{hyperref}
\usepackage{braket}
\usepackage{mathbbol}
\usepackage{graphicx}
\usepackage{dcolumn}
\usepackage{bm}
\usepackage{soul}

\usepackage{xcolor}

\begin{document}

\preprint{APS/123-QED}

\title{Collisional interferometry of Levitons in quantum Hall edge channels at $\nu=2$}

\author{Giacomo Rebora}
 \affiliation{Dipartimento di Fisica, Universit\`a di Genova, Via Dodecaneso 33, 16146, Genova, Italy}
 \affiliation{CNR-SPIN, Via Dodecaneso 33, 16146, Genova, Italy}
\author{Matteo Acciai}
\affiliation{Department of Microtechnology and Nanoscience (MC2), Chalmers University of Technology, S-412 96
G\"oteborg, Sweden}
\affiliation{Dipartimento di Fisica, Universit\`a di Genova, Via Dodecaneso 33, 16146, Genova, Italy}
\author{Dario Ferraro}%
\affiliation{Dipartimento di Fisica, Universit\`a di Genova, Via Dodecaneso 33, 16146, Genova, Italy}%
\affiliation{CNR-SPIN, Via Dodecaneso 33, 16146, Genova, Italy}
\author{Maura Sassetti}
\affiliation{Dipartimento di Fisica, Universit\`a di Genova, Via Dodecaneso 33, 16146, Genova, Italy}%
\affiliation{CNR-SPIN, Via Dodecaneso 33, 16146, Genova, Italy}

\date{\today}

\begin{abstract}
We consider a Hong-Ou-Mandel interferometer for Lorentzian voltage pulses applied to Quantum Hall edge channels at filling factor $\nu=2$. Due to inter-edge interactions, the injected electronic wave-packets fractionalize before partitioning at a quantum point contact. Remarkably enough, differently from what theoretically predicted and experimentally observed by using other injection techniques, we demonstrate that, when the injection occurs through time-dependent voltage pulses (arbitrarily shaped), the Hong-Ou-Mandel noise signal always vanishes for a symmetric device, and that a mismatch in the distances between the injectors and the point of collision is needed in order to reduce the visibility of the dip. We also show that, by properly tuning these distances or by applying different voltages on the two edge channels in each arm of the interferometer, it is possible to estimate the intensity of the inter-edge interaction. The voltage pulses are chosen of the Lorentzian type because of their experimental relevance.
\end{abstract}

\maketitle


\section{\label{sec:introd} Introduction}
The progresses in the experimental control of individual electronic degrees of freedom ballistically propagating in mesoscopic devices led to the birth of a new branch of condensed matter physics known as Electron Quantum Optics (EQO)~\cite{Grenier11,Bocquillon14,Bauerle18}.
In this framework, intensity interferometers such as the Hanbury-Brown-Twiss (HBT)~\cite{Hanbury56} and the Hong-Ou-Mandel (HOM)~\cite{Hong87} have been realized by partitioning electronic wave-packets~\cite{Bocquillon12} and making them collide with a tunable delay~\cite{Bocquillon13b} at a Quantum Point Contact (QPC). These seminal experiments realized with voltage-driven mesocopic capacitors~\cite{Feve07,Mahe10,Parmentier12}, first proposed theoretically in Refs.~\cite{Buttiker93, Ol'khovskaya08,Moskalets11,Jonckheere12,Haack13}, have been realized by means of periodic trains of electrons and holes with wave-packets peaked in energy.

One of the main differences between the photonic and the electronic case is represented by the fact that electrons are charged interacting particles. This leads to many-body effects which strongly affect the dynamics of excitations and play a major role in various experimental situations. In particular this is true when experiments are carried out in Quantum Hall (QH) edge channels at filling factor $\nu=2$, where inter-channel interaction cannot be neglected~\cite{Bocquillon13}. This emerges dramatically in HOM experiments realized with a driven mesoscopic capacitor in the non-adiabatic regime~\cite{Ol'khovskaya08,Jonckheere12,Moskalets13,Dashti19}, where the visibility of the predicted dip in the auto-correlated noise as a function of the injection delay~\cite{Ol'khovskaya08,Jonckheere12}, signature of the anti-bunching of electrons, is strongly reduced due to electron-electron interactions~\cite{Wahl14,Freulon15,Marguerite16}.

An alternative protocol for the injection of electrons consists in the application of a train of well designed time-dependent voltage pulses~\cite{Misiorny18}. According to what has been discussed by Levitov and coworkers~\cite{Levitov96,Ivanov97,Keeling06}, a properly quantized Lorentzian drive leads, in a non-interacting system, to the injection of purely electronic wave-packets without any additional electron-hole pair contribution. This prediction has been validated experimentally~\cite{Dubois13,Glattli16} through the realization of HBT and HOM collisional experiments in non-interacting narrow constrictions realized in two-dimensional electron gases. These low-energy excitations, usually called Levitons, are predicted to be robust with respect to interaction-induced decoherence~\cite{Ferraro14} and anomalous correlations among electrons~\cite{Acciai19, Ronetti20}. Moreover, this robustness survives also in very strong interacting environments such as Fractional QH states~\cite{Keeling06,Safi10,Rech17,Vannucci17}. Here, remarkable features related to a crystallization of Levitons in the time domain have been reported by some of the authors~\cite{Ronetti18,Ferraro18b,Ronetti19} and can be observed in HOM interferometers.%

As stated above, EQO in QH edge states at filling factor $\nu=2$ has been investigated so far in the case of the emitted excitations generated via driven mesoscopic capacitors (as experiments in this regime typically involve this kind of source)~\cite{Bocquillon13,Wahl14,Freulon15,Marguerite16,Cabart18}. Some theoretical works have also addressed the case of injection at $\nu=2$ via voltage pulses, focusing on the evolution of excitations due to interactions on the HBT noise signal~\cite{Grenier13,Ferraro14,Acciai18,Cabart18}.
Thus, a detailed theoretical analysis of collisional HOM setups for voltage pulses and in particular for Levitons in QH edge channels at $\nu=2$, even if relevant for the interpretation of forthcoming experiments, is still missing. This paper intends to fill this gap by studying the signatures of inter-edge interactions in the profile of the HOM noise signal emphasizing the difference with the mesoscopic capacitor set-up. We demonstrate that the visibility of the central dip is always maximal (the dip goes to zero) when the excitations are injected through time-dependent voltages of arbitrary shape and the setup is symmetric, namely the distance of the two injectors from the QPC is the same for both sides of the interferometer. This can be seen as a signature of the robustness of voltage signals against decoherence, differently from what observed for the driven mesoscopic capacitor~\cite{Bocquillon13,Freulon15,Marguerite16}. Our results will be discussed in detail for the particular case of Lorentzian voltage pulses due to their great relevance from the experimental point of view and in a single electron source perspective.

We will also observe that the visibility can be strongly reduced only in an asymmetric device, where a mismatch in the lengths of the two arms of the interferometer is present. Moreover, from the evolution of the visibility as a function this mismatch it is also possible to extract information about the strength of the electron-electron interaction. We will also propose a more direct measurement of the interaction based on the fact that, when properly tuned voltages are applied on both the edge channels of the two arms of device, it is possible to cancel all the interaction-dependent features of the HOM noise recovering the non interacting case. Such kind of fine tuning allows to deduce the value of the interaction between the channels. 

The paper is organized as follows. Section~\ref{sec:model} describes the model for two interacting QH edge channels in terms of the edge-magnetoplasmon scattering matrix formalism. In Section~\ref{sec:noise} we describe the general aspects of the HOM interferometry for electrons injected by means of voltage drives.
In Section~\ref{sec:results} we focus on the injection of Levitons discussing the essential features of the HOM noise as a function of the delay in the injection for both a symmetric and an asymmetric set-up. In Section~\ref{sec:meas} we demonstrate that a collisional HOM experiment allows to measure the inter-edge interactions by studying the evolution of side dips in the HOM signal when the two edge channels are driven independently. Finally, Section~\ref{sec:concl} is devoted to the conclusions. Technical details of the calculations are reported in three Appendices. 

\begin{figure}[b]
\includegraphics[width=\columnwidth]{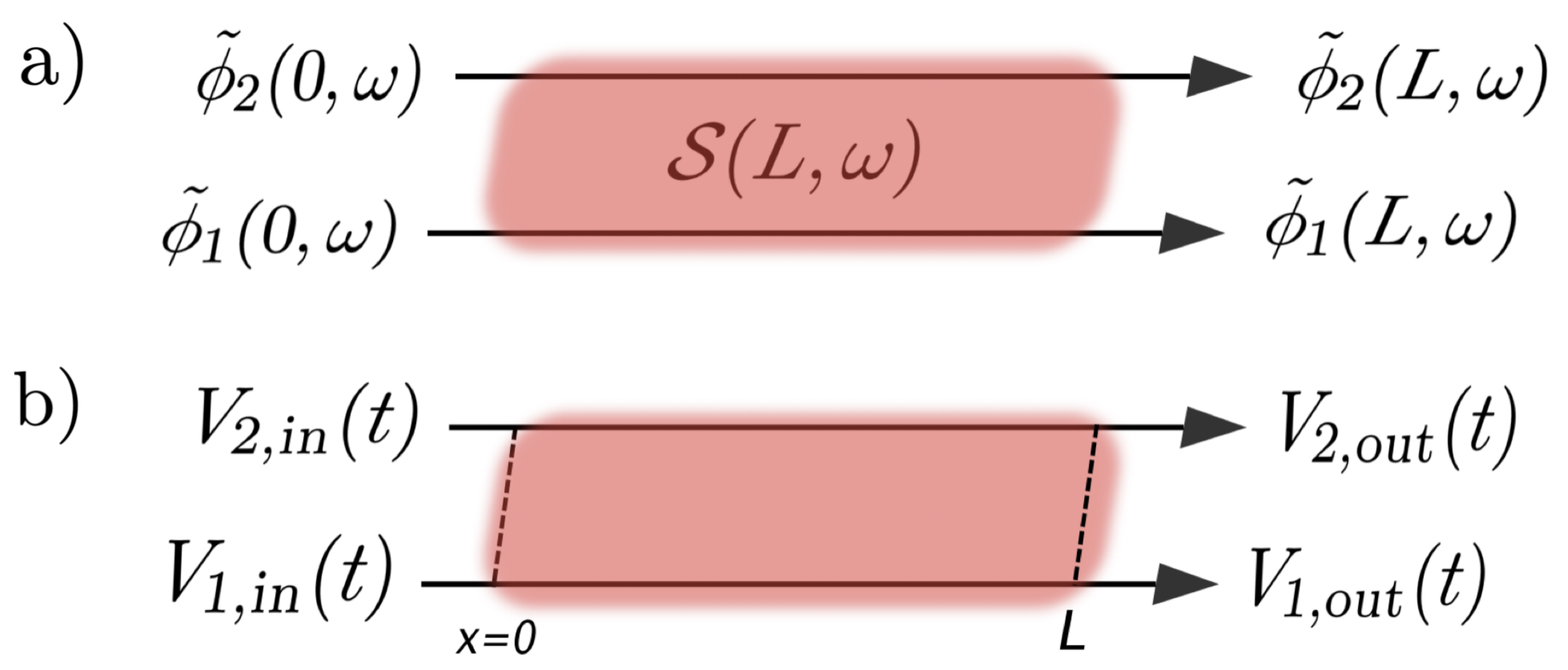}
\caption{\label{fig:fig1} Schematic view of QH channels at integer filling factor $\nu=2$. The shaded red area represents the interaction region, which has a finite length $L$ and is described by the scattering matrix $\mathcal{S}(L,\omega)$. a) After passing this region, the incoming bosonic fields $\tilde{\phi}_{1,2}(0,\omega)$ are transformed into the outgoing ones $\tilde{\phi}_{1,2}(L,\omega)$. b) The input voltages $V_{1/2}^{\mathrm{in}}$, applied to the edge channels. Due to interactions, the excitations emerging after the propagation from $x=0$ to $x=L$ are equivalent to those that would be generated by the output voltages $V_{1/2}^{\mathrm{out}}$ applied to the channels directly at the end of the interaction region. These output voltages are related to the incoming ones by Eq.~\eqref{eq:voltage}.}
\end{figure}

\section{\label{sec:model} Model}
We consider a QH bar at filling factor $\nu=2$. The two copropagating edge channels are assumed to interact along a region of finite length $L$ via a screened ($\delta$-like) Coulomb repulsion~\cite{Sukhorukov07, Levkivskyi08, Degiovanni10, Wahl14, Ferraro14, Ferraro17} which correctly reproduces the experimental observations at low enough energies~\cite{Bocquillon13,Hashisaka17,Hashisaka18}. This mechanism is well described within the chiral Luttinger liquid theory based on bosonic collective excitations called edge-magnetoplasmons \cite{Degiovanni10, Ferraro14, Ferraro17,Acciai18}. Here, the Hamiltonian density $\mathcal{H}$ describing the two copropagating channels along each edge is the sum of a kinetic term $\mathcal{H}_0$ and an interaction contribution $\mathcal{H}_\mathrm{int}$. Following Wen's hydrodynamical model~\cite{Wen95} they are written as ($\hbar=1$)
\begin{align}
    \mathcal{H}_0&=\sum_{i=1,2}\frac{v_i}{4\pi}\left( \partial_x\phi_{i}(x)\right)^2 \label{eq:0}\\
    \mathcal{H}_{\mathrm{int}}&=\frac{u}{2\pi}\left( \partial_x\phi_{1}(x)\right)\left( \partial_x\phi_{2}(x)\right)\,,\label{eq:int}
\end{align}
where the index $i=1,2$ labels inner and outer channels respectively (see Fig.~\ref{fig:fig1}a), while $\phi_{i}$ are chiral bosonic fields satisfying the commutation relations
\begin{equation}
[\phi_i(x),\phi_j(y)]=i \pi \,\mathrm{sign}(x-y) \delta_{ij}
\end{equation} 
and related to the particle density operator by~\cite{Miranda03}
\begin{equation}
\rho_{i}(x)=\frac{1}{2\pi}\partial_x \phi_{i}(x)
\end{equation} 
where the fermionic field $\psi_i(x)$ is related to the bosonic one $\phi_i(x)$ via
\begin{equation}
    \psi_i(x) = \frac{\mathcal{F}_i}{\sqrt{2\pi a}}e^{-i\phi_i(x)}
\end{equation}
where $a$ is a short distance cut-off and $\mathcal{F}_i$ are the Klein factors~\cite{Wen95, Miranda03}.

In Eqs~(\ref{eq:0})-(\ref{eq:int}) $v_i$ are the bare propagation velocities of the two edge channels (here, without loss of generality, we assume $v_{1}\geq v_{2}$) and $u$ is the intensity of the inter-edge coupling. The full interacting problem can be diagonalized through a rotation in the bosonic fields space by an angle $\theta$ satisfying
\begin{equation}
\tan(2\theta)=\frac{2u}{(v_1-v_2)}. 
\end{equation} 
This parameter encodes the interaction strength, $\theta=0$ being the non-interacting limit and $\theta=\pi/4$ representing what in the literature is usually indicated as the strong interacting regime~\cite{Bocquillon13,Wahl14,Levkivskyi08}. However, the stability of the model, namely the request that both eigenvelocities are positive~\cite{Braggio12}, imposes a constraint on the maximum admissible value of $u$ (see below), therefore strictly speaking this limit can be properly obtained only for $v_1=v_2$ by keeping $u$ fixed~\cite{Kovrizhin12}. Experimentally, values of $\theta$ ranging from $\theta\approx \pi/6$~\cite{Inoue14,Rodriguez20} to $\theta\approx\pi/4$ \cite{Bocquillon13} have been reported, indicating that this parameter strongly depends on the specific details of the considered set-ups.   

The rotation leads to two new bosonic fields, defined by
\begin{align}
    \phi_\rho(x)&=\cos\theta\,\phi_1(x)+\sin\theta\,\phi_2(x)\\
    \phi_\sigma(x)&=-\sin\theta\,\phi_1(x)+\cos\theta\,\phi_2(x)\,,
\end{align}
in terms of which the full diagonalized Hamiltonian density becomes
\begin{equation}
    \mathcal{H}=\sum_{\beta=\rho,\sigma}\frac{v_\beta}{4\pi}(\partial_x\phi_\beta(x))^2\,.
\end{equation}
These fields are associated with two new collective modes: a slow dipolar and a fast charge mode propagating respectively with velocities $v_\sigma$ and $v_\rho$, where 
\begin{equation}
    v_{\rho/\sigma}=\left(\frac{v_1+v_2}{2}\right)\pm\frac{1}{\cos(2\theta)}\left(\frac{v_1-v_2}{2}\right).
\end{equation}

The dynamics of the edge channels can be solved within a scattering formalism~\cite{Degiovanni10, Sukhorukov15}. As depicted in Fig.~\ref{fig:fig1}a, the fields outgoing from a scattering region of finite length $L$ are related to the incoming ones through the edge-magnetoplasmon scattering matrix $\mathcal{S}(L,\omega)$ as
\begin{equation}
    \begin{pmatrix}
    \tilde{\phi}_{1}(L,\omega) \\
    \tilde{\phi}_{2}(L,\omega)
    \end{pmatrix}=\mathcal{S}(L,\omega)\begin{pmatrix}
    \tilde{\phi}_{1}(0,\omega) \\
    \tilde{\phi}_{2}(0,\omega)
    \end{pmatrix}.
    \label{eq:phi-out}
\end{equation}
Here, $\tilde\phi_{1/2}(x,\omega)$ is the Fourier transform, with respect to time, of $\phi_{1/2}(x,t)$ and \cite{Degiovanni10, Ferraro17}
\begin{equation}
\mathcal{S}=
\begin{pmatrix}
\cos^2{\theta}\,e^{i\omega\tau_\rho}+\sin^2{\theta}\,e^{i\omega\tau_\sigma} & \sin{\theta}\cos{\theta}(e^{i\omega\tau_\rho}-e^{i\omega\tau_\sigma}) \\
\sin{\theta}\cos{\theta}(e^{i\omega\tau_\rho}-e^{i\omega\tau_\sigma}) & \sin^2{\theta}\,e^{i\omega\tau_\rho}+\cos^2{\theta}\,e^{i\omega\tau_\sigma}
 \end{pmatrix}
 \label{eq:smatrix}
\end{equation}
where $\tau_{\rho/\sigma}=L/v_{\rho/\sigma}$ are the times of flight associated with fast and slow modes respectively.

Following Refs.~\cite{Dubois13,Dubois13b,Glattli16} we can consider an electron source modeled as an ohmic contact coupling each channel to a time-dependent voltage source and allowing us to control the injection of electrons through voltages $V_{1,\mathrm{in}}(t)$ and $V_{2,\mathrm{in}}(t)$ applied to the inner and the outer channel respectively, according to the conventional coupling Hamiltonian
\begin{equation}
\mathcal{H}_{U}=-e\int \rho_{i}(x)U_{i,\mathrm{in}}(x,t) dx\,,
\end{equation}
where $i=1\,(2)$ labels the inner (outer) channel and $-e$ $(e>0)$ is the electron charge and $U_{i,\mathrm{in}}(x,t)$ describes the effect of the voltage source connected to the channels. We write it as $U_{i,\mathrm{in}}(x,t)=\Theta(-x)V_{i,\mathrm{in}}(t)$ where $V_{i,\mathrm{in}}(t)$ is the time-dependent voltage of the source and the Heaviside step function $\Theta(-x)$ specifies the region where this potential is applied~\cite{Rech17, Vannucci17}.

This classical potential, coupled to the charge density along the edge according to the above Equation can then be seen as an external classical forcing for a quantum harmonic oscillator leading to the generation of a coherent state of the edge-magnetoplasmons along the edge channels. The displacement parameter associated to this coherent state is derived by solving the equations of motion for the bosonic fields (considering the complete Hamiltonian $\mathcal{H}+\mathcal{H}_{U}$) and is proportional to the Fourier transform of the voltages $\tilde{V}_{i,\mathrm{in}}(\omega)$~\cite{Safi99, Grenier13}. In the frequency domain the interacting region acts as a beam-splitter for this coherent state through the edge-magnetoplasmon $\mathcal{S}$ in exactly the same way as for the bosonic modes in absence of voltage, namely
\begin{equation}
    \begin{pmatrix}
    \tilde{V}_{1, \mathrm{out}}(\omega) \\
    \tilde{V}_{2, \mathrm{out}}(\omega)
    \end{pmatrix}=\mathcal{S}(L,\omega)\begin{pmatrix}
    \tilde{V}_{1, \mathrm{in}}(\omega) \\
    \tilde{V}_{2, \mathrm{in}}(\omega)
    \end{pmatrix}.
    \label{eq:phi-out}
\end{equation}
In the time domain (see Fig.~\ref{fig:fig1}b) this leads to
\begin{equation}
 \begin{split}
      V_{1,\mathrm{out}}(t)=&\cos^2{\theta}\,V_{1,\mathrm{in}}(t-\tau_\rho)+\sin^2{\theta}\,V_{1,\mathrm{in}}(t-\tau_\sigma)\\ +&\sin{\theta}\cos{\theta}[V_{2,\mathrm{in}}(t-\tau_\rho)-V_{2,\mathrm{in}}(t-\tau_\sigma)]\\
      V_{2,\mathrm{out}}(t)=&\sin{\theta}\cos{\theta}[V_{1,\mathrm{in}}(t-\tau_\rho)-V_{1,\mathrm{in}}(t-\tau_\sigma)]\\ +&\sin^2{\theta}\,V_{2,\mathrm{in}}(t-\tau_\rho)+\cos^2{\theta}\,V_{2,\mathrm{in}}(t-\tau_\sigma)
 \end{split}
 \label{eq:voltage}
\end{equation}
clearly showing that, at the end of the interaction region, the two incoming voltages are mixed.

\begin{figure}[t]
\includegraphics[width=\columnwidth]{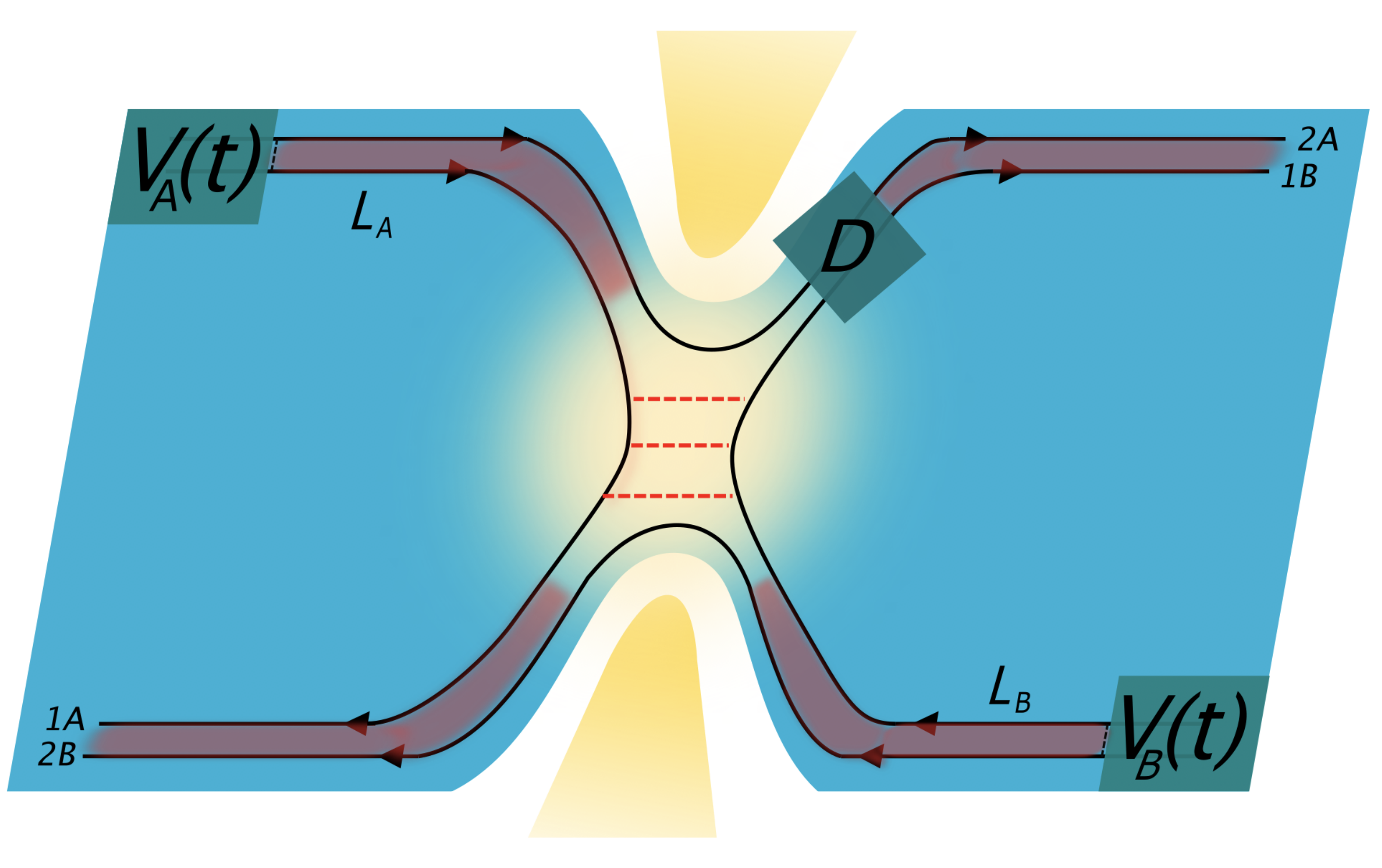}
\caption{\label{fig:fig2} Schematic view of the HOM interferometer. Two pairs of copropagating and interacting edge states, on opposite sides of a QH bar, meet at a QPC. The electrons sources $A$ and $B$ are modeled as ohmic contacts which are used to drive each edge with time-dependent voltages (here $V_{A/B}(t)$ is a compact notation to indicate $V_{1/2,\mathrm{in}}^{A/B}(t)$, which are the voltages shown in Fig.~\ref{fig:fig1}b and that are mixed by the interaction region indicated with the shaded red area). A detector $D$ is placed just after the QPC in order to measure current correlations. Notice that the region of the QPC is brighter to indicate the fact that here the electron-electron interaction is screened.}
\end{figure}
\section{\label{sec:noise} General aspects of HOM interferometry}
We now consider the effect of interaction in a HOM experiment where electronic wave-packets, generated by means of applied voltage pulses, collide at a QPC with a controlled delay in time. Unlike the injection with a driven mesoscopic capacitor~\cite{Feve07,Bocquillon13b, Marguerite16}, this case still lacks of a detailed investigation from both the theoretical and the experimental point of view.  

Fig.\ \ref{fig:fig2} shows the HOM interferometer. Here, excitations emitted by the voltage sources $A$ and $B$ fractionalize when going through the interacting regions and are then partitioned at a QPC. For the moment, we assume that the injection only occurs into the inner channels of each edge, postponing the analysis of a more general case to Sec.\ \ref{sec:meas}. Therefore we set $V_{2,\mathrm{in}}^{A/B}(t)=0$, where the notation now takes into account the fact that one can apply a voltage both to the $A$ and the $B$ source. It is worth noting that, as far as QH edge states in the integer regime are concerned, the different edge channels can be addressed independently by means of additional upstream QPCs~\cite{Altimiras10,leSueur10} or quantum dots with high transparency~\cite{Bocquillon13b}. Moreover, we assume that the partitioning at the QPC involves the inner channels only, which we label as $1A$ (right-moving) and $1B$ (left-moving), related to the incoming fermionic fields $\psi_{1A}^{\mathrm{I}}$ and $\psi_{1B}^{\mathrm{I}}$ that are evaluated immediately before the QPC. Such a situation can be implemented by properly tuning the QPC transparency in such a way that the outer channels are completely transmitted, while the inner ones are also partially reflected~\cite{Bocquillon13b, Wahl14, Marguerite16}. Thus, the brighter region of the QPC (see Fig.~\ref{fig:fig2}) is not included in the interacting region and fermions are locally free at this location. According to this and assuming a local tunneling, the free fermionic fields, outgoing from the QPC, are related to the incoming ones through a scattering matrix
\begin{equation}
    \begin{pmatrix}
    \psi_{1A}(t) \\
    \psi_{1B}(t)
    \end{pmatrix}_{\mathrm{O}}= \begin{pmatrix} \sqrt{R} & i\sqrt{T} \\ i\sqrt{T} & \sqrt{R}\end{pmatrix}
    \begin{pmatrix}
    \psi_{1A}(t) \\
    \psi_{1B}(t)
    \end{pmatrix}_{\mathrm{I}}
    \label{eq:matr}
\end{equation}
where $T$ and $R=1-T$ are positive real parameters describing the probability for a particle to be transmitted or reflected, respectively. These probabilities are assumed as energy independent, a condition which is typically well fulfilled in experiments~\cite{Bocquillon12, Bocquillon13b, Marguerite16}. This scattering approach for fermionic fields is justified, in our specific case of interacting channels at $\nu=2$, as long as both the inter-edge interaction and the tunneling are local (see for example Supplementary Material of Ref.~\cite{Wahl14}). According to the chirality and locality of the coupling we can consider the interaction region extending from just after the injection point to just before the QPC~\cite{Degiovanni10,Ferraro14,Ferraro17}. This mathematical description is physically motivated by the fact that both the contacts used to apply the voltage and the gates that realize the QPC locally enhance the screening of the interaction, that can be therefore assumed as negligible in these two regions. This theoretical approach already showed a very good agreement with the experimental observation for HOM interferometers realized using driven mesoscopic capacitors as single electron sources~\cite{Bocquillon13,Marguerite16}.

Following what is usually investigated in EQO experiments, we focus our attention on the zero-frequency auto-correlated noise $S_\mathrm{HOM}$, which we evaluate just after the QPC. This quantity is defined as~\cite{Blanter00,Martin05,Ferraro14b}
\begin{equation}
S_{\mathrm{HOM}}=\int \left[ \braket{I_{D}(t)I_{D}(t')}-\braket{I_D(t)}\braket{I_D(t')} \right] dt dt', 
\label{eq:noise-def}
\end{equation}
where $I_D(t)$ is the total current arriving at the detector $D$ (see Fig.~\ref{fig:fig2}) and it is composed of the currents flowing in the channels $2A$ and $1B$:
\begin{equation}
    I_{D}(t)=I_{2A}(t)+I_{1B}(t)\,.
\end{equation}
The current operator on a given channel $j=2A,1B$ reads $I_j(t)=-ev_F:\psi^{\dagger}_j(t)\psi_j(t):$, where $:\dots:$ denotes the normal ordering with respect to the Fermi sea and fermionic fields are evaluated at the level of the detector $D$.

In full generality, the HOM noise can be expressed as
\begin{equation}
S_\mathrm{HOM}=S_{2A,2A}+S_{2A,1B}+S_{1B,2A}+S_{1B,1B}\,,
\label{eq:son}
\end{equation}
where ($i,j=2A, 1B$)
\begin{equation}
    S_{ij}=\int \left[ \braket{I_i(t)I_{j}(t')}-\braket{I_i(t)}\braket{I_j(t')} \right] dt dt'\,.
\end{equation}
 
The notation $S_\mathrm{HOM}$ is chosen to emphasize that we are dealing with the zero-frequency noise in the HOM configuration, i.e.\ when both sources are on. We note that in Eq.~\eqref{eq:son} the first contribution $S_{2A,2A}$ consists only of the current auto-correlations of the totally transmitted external channel but this does not affect the measurements because its contribution is zero. Also the terms $S_{2A,1B}$ and $S_{1B,2A}$ do not contribute, due to the fact that averages involving current operators in different channels factorize because no interaction occurs at the level of the QPC. Therefore, the only relevant contribution in Eq.~\eqref{eq:son} is $S_{1B,1B}$ which involves terms referring to both inner channels. This is because, according to Eq.~\eqref{eq:matr}, the fermionic field $\psi^{\mathrm{O}}_{1B}$ at the output of the QPC is expressed in terms of both incoming fields $\psi^{\mathrm{I}}_{1A}$ and $\psi^{\mathrm{I}}_{1B}$.
In order to simplify the notation, in the following discussion we will refer to the inner channels $1A$ and $1B$ just as $A$ and $B$.

By using Eqs.~\eqref{eq:matr} and \eqref{eq:noise-def} we can express the total noise $S_\mathrm{HOM}$ as \cite{Ferraro13}
\begin{equation}
    S_\mathrm{HOM}=-(e v_F)^2 RT\int  \Delta Q(t,t')dt dt'
    \label{eq:noise}
\end{equation}
where
\begin{equation}
\begin{split}
    \Delta Q(t,t')=&\Delta \mathcal{G}_{A}^{(e)}(t',t) \Delta \mathcal{G}_B^{(h)}(t',t) +\Delta \mathcal{G}_{A}^{(h)}(t',t) \Delta \mathcal{G}_B^{(e)}(t',t)\\
    +&\Delta \mathcal{G}_{A}^{(e)}(t',t) \mathcal{G}_{F,B}^{(h)}(t',t)+ \Delta \mathcal{G}_{A}^{(h)}(t',t) \mathcal{G}_{F,B}^{(e)}(t',t)\\
     +& \mathcal{G}_{F,A}^{(e)}(t',t) \Delta\mathcal{G}_{B}^{(h)}(t',t)+ \mathcal{G}_{F,A}^{(h)}(t',t) \Delta\mathcal{G}_{B}^{(e)}(t',t).
    \label{eq:G}
    \end{split}
\end{equation}
In Eq.~\eqref{eq:G} $\Delta \mathcal{G}_{A/B}^{(e/h)}$ are the non-equilibrium excess first order coherence  functions~\cite{Grenier112, Haack13, Moskalets16}
\begin{equation}
    \Delta \mathcal{G}^{(e/h)}_{A/B}(t',t)=\mathcal{G}^{(e/h)}_{A/B}(t',t)-\mathcal{G}^{(e/h)}_{F,A/B}(t'-t)\,,
    \label{eq:vi}
\end{equation}
where
\begin{subequations}
\begin{align}
    \mathcal{G}_{A/B}^{(e)}(t',t)&=\Braket{\psi_{A/B}^\dagger(t)\psi_{A/B}(t')}\\
    \mathcal{G}_{A/B}^{(h)}(t',t)&=\Braket{\psi_{A/B}(t)\psi_{A/B}^\dagger(t')}
\end{align}
\end{subequations}
are correlators evaluated over the non-equilibrium state induced by the voltage injection, whereas $\mathcal{G}_{F,A/B}^{(e/h)}$ are the correlation functions for the equilibrium states (i.e.\ when no drive is applied) and are evaluated over the Fermi sea. The channel label $(A/B)$ will be dropped in the following when referring to the equilibrium correlation functions, as they are assumed identical for both channels.
The effect of the external voltage drive can be properly taken into account with a phase factor~\cite{Ferraro13,Grenier13}, in such a way that Eq.\ \eqref{eq:vi} is rewritten as
\begin{equation}
    \Delta \mathcal{G}^{(e/h)}_{A/B}(t',t)=\mathcal{G}^{(e/h)}_{F}(t'-t)\left( e^{\mp i\varphi_{A/B}(t,t')}-1\right)
    \label{eq:phase}
\end{equation}
where
\begin{equation}
\varphi_{A/B}(t,t')=e\int_{t'}^{t}V_{1,\mathrm{out}}^{A/B}(\tau)d\tau   
\label{varphi}
\end{equation} 
is the phase contribution due to the time dependent voltage, carrying information about interaction effects according to Eq.~\eqref{eq:voltage}. 
It is worth noting that, limited to the injection through voltage and under the assumption of local interaction acting over a finite length, the coherence functions can be written as the free fermionic ones times phase factors encoding the effect of the applied voltage and the interaction. Therefore, our system can be mapped onto a free fermion problem subject to a modified voltage which takes into account the fractionalization effects~\cite{Grenier13}. Incidentally this fact can be seen as a further validation of Eq.~(\ref{eq:matr}).

By replacing Eq.\ \eqref{eq:phase} into Eq.\ \eqref{eq:G}, the correlation function $\Delta Q(t,t')$ can be expressed as
\begin{equation}
\begin{split}
    \Delta Q(t,t')&=2\mathcal{G}^{(e)}_{F}(t'-t)\mathcal{G}^{(h)}_{F}(t'-t)\\
    &\quad\times\big[1-\cos(\varphi_{A}(t,t')-\varphi_B(t,t'))\big]\,.
 \end{split}
 \label{eq:new}
\end{equation}
If one of the two sources is switched off, the above formula simplifies and the HBT noise associated with the partitioning of excitations incoming only in one arm of the interferometer is recovered $(i=A,B)$~\cite{Ferraro13}:
\begin{equation}
\begin{split}
S_{\mathrm{HBT},i}=-2(ev_F)^2 RT\int & dt\,dt' \mathcal{G}^{(e)}_{F}(t'-t)\mathcal{G}^{(h)}_{F}(t'-t)\\
&\times\left[1-\cos(\varphi_i(t,t'))\right]\,.
\end{split}
\end{equation}

In the following, according to what is usually done in conventional HOM experiments with voltage pulses~\cite{Dubois13,Glattli16}, we consider the two sources $A$ and $B$ to be driven by identical signals apart from a controlled time delay $\delta$, namely \begin{equation}
V_{1,\mathrm{in}}^{B}(t)=V_{1,\mathrm{in}}^{A}(t+\delta).
\label{eq:relation-va-vb}
\end{equation} 
An important consequence arises when we consider the interaction strengths and the distances between the sources and the QPC to be equal in both arms of the interferometer (symmetric configuration with $\theta^{A}=\theta^{B}=\theta$ and $L_{A}=L_{B}=L$). In this case, the voltages $V_{1,\mathrm{out}}^{A/B}$ after the interacting regions are the same for both arms. This can be easily seen from Eq.~\eqref{eq:voltage} where it is clear how these voltages depend on the interaction strength $\theta$ and on the interaction length $L$ (via the times of flight $\tau_{\rho/\sigma}$). As a result, at zero injection delay $\delta=0$ one has $\varphi_A(t,t')=\varphi_B(t,t')$, leading to $\Delta Q(t,t')=0$. Therefore we arrive at the consequence that, even in the presence of interactions, the HOM noise in a symmetric configuration \emph{always} vanishes for a synchronized emission in the two incoming channels ($\delta=0$), \emph{regardless} of the particular form of the signal used for the time-dependent voltage injection. Notice that these considerations still hold also in the case of a long-range interaction~\cite{Grenier13, Cabart18, Freulon15} as long as it preserves the symmetry of the set-up.

The injection via the mesoscopic capacitor occurs at a well defined energy above the Fermi level and it has been shown~\cite{Ferraro14} that in this case the emitted wave-packets undergo a relaxation towards low-energy degrees of freedom before the process of fractionalization takes place. On the contrary, voltage-generated excitations are robust in this respect, as the energy relaxation does not occur for them~\cite{Ferraro14,Cabart18} and they are only affected by the fractionalization process during their propagation through the interacting region.

This qualitative difference is consistent with our results, showing that the excitations injected via voltage pulses are robust and do not display any suppression of the HOM dip at zero delay.

We recall that a standard experimental procedure consists in normalizing the measured HOM signal with respect to the HBT ones~\cite{Bocquillon13}, thus defining the ratio
\begin{equation}
\mathcal{R}(\delta)=\frac{S_{\mathrm{HOM}}(\delta)}{S_{\mathrm{HBT},A}+S_{\mathrm{HBT},B}},
\label{eq:ratio}
\end{equation}
where we have taken into account the fact that the HOM noise contribution is the only one which depends on the time delay $\delta$.
The noise in Eq.~\eqref{eq:noise} can be rewritten in terms of the average time $\bar{t}=(t+t')/2$ and of the time difference $\tau=t-t'$ as (adapting the definition to the case of a periodic drive \cite{Dubois13b})
\begin{equation}
 S_\mathrm{HOM}=-(e v_F)^2 RT\! \int_{-\frac{\mathcal{T}}{2}}^{\frac{\mathcal{T}}{2}} \frac{d\bar{t}}{\mathcal{T}}\int_{-\infty}^{+\infty}\!d\tau \Delta Q\left(\bar{t}+\frac{\tau}{2},\bar{t}-\frac{\tau}{2}\right).
    \label{eq:newnoise}
\end{equation}
These integrals are performed analytically in Appendix \ref{app1} by introducing the Fourier series
\begin{equation}
    e^{-ie\int_0^tV(t')dt'}=e^{-iq\Omega t}\sum_{l=-\infty}^{+\infty}\,p_l(q)\,e^{-il\Omega t}
    \label{eq:four}
\end{equation}
where $\Omega=2\pi/\mathcal{T}$ and the photoassisted coefficients $p_l$ are linked to the probability amplitude for photon absorption ($l>0$) or emission ($l<0$)~\cite{Dubois13}.

By using this approach, the ratio \eqref{eq:ratio} can be written as
\begin{equation}
    \mathcal{R}(\delta)=\sum_{l=-\infty}^{+\infty}\frac{{|\mathcal{P}_l(q;\delta)|}^2|\Omega l|}{{|\widetilde{p}_{l,A}(q)|}^2|\Omega (l+q)|+{|\widetilde{p}_{l,B}(q)|}^2|\Omega (l+q)|},
    \label{eq:rapp}
\end{equation}
where $\widetilde{p}_{l,A/B}(q)$ and $\mathcal{P}_l(q;\delta)$ are new photoassisted coefficients defined in Appendix~\ref{app1} (Eqs.~\eqref{eq:coeff2} and~\eqref{eq:coeffp}) and they can be expressed as functions of amplitudes $p_l$ defined in Eq.~\eqref{eq:four}. They are related to the phases $\varphi_A-\varphi_B$ and $\varphi_{A/B}$, respectively, and fully take into account the effects of interactions. It is worth noting that Eq.~\eqref{eq:rapp} as well as all the following results are obtained in the zero temperature limit, thermal corrections being marginal in realistic experimental conditions~\cite{Bocquillon13,Dubois13,Marguerite16,Glattli16}.

In the next Section, we specify the above general analysis to the case of Lorentzian pulses, a particularly relevant drive in the context of EQO \cite{Keeling06,Dubois13}, considering symmetric and asymmetric configurations. Both of them are analyzed by relying on the general expression \eqref{eq:rapp}, where the proper photoassisted coefficients of Lorentzian pulses [see Eq.~\eqref{eq:coefflor}] will be used.


\section{\label{sec:results} HOM interferometry for Levitons}

In the previous Section, we have proved that the excess noise in a symmetric HOM configuration, is always zero for simultaneous injection from the the sources, independently of the shape of the voltage and of the interactions occurring along the channels. In this Section we will focus on a specific form for the voltage drive which is particularly relevant in the context of experimental EQO and we present the results for the ratio $\mathcal{R}$ as a function of the time delay $\delta$ between the two sources. In order to properly describe realistic experimental configurations, we consider the two sources to be periodically driven in time. The injection of a periodic train of single electrons, without hole contributions, is possible by applying properly quantized Lorentzian voltage pulses~\cite{Levitov96,Keeling06,Dubois13} of the form
\begin{equation}
    V_{1,\mathrm{in}}^A(t)\equiv V(t)=V_0\sum_{j\in \mathbb{Z}} \frac{\tau_0}{\tau_0^2+(t-j\mathcal{T})^2}\,,
    \label{eq:lor}
\end{equation}
where $V_0=-2 q/e$ ($\hbar=1$), with $q\in\mathbb{N}$. When $q=1$, one electron per period $\mathcal{T}$ is emitted, realizing a train of so-called Levitons~\cite{Dubois13,Glattli16}. In Eq.~\eqref{eq:lor} $\tau_0$ represents the width in time of each Lorentzian pulse of the periodic train. 


\subsection{\label{sec:symmetric}Symmetric setup}
In this Section we analyze what happens to the noise ratio $\mathcal{R}$, in Eq.~\eqref{eq:rapp}, when identical Lorentzian voltage pulses with unitary charge ($q=1$) are applied to both contacts. We consider a symmetrical configuration for the interferometer, meaning that the lengths of the two interacting regions are equal ($L_{A}=L_{B}=L$), as well as the inter-edge interaction strength ($\theta^{A}=\theta^{B}=\theta$) in the two incoming channels. It is worth noticing that in this situation the photoassisted coefficients $\widetilde{p}_{l,A}(q)$ and $\widetilde{p}_{l,B}(q)$ entering in Eq.~\eqref{eq:rapp} are equal.

\begin{figure}[t]
\includegraphics[width=\columnwidth]{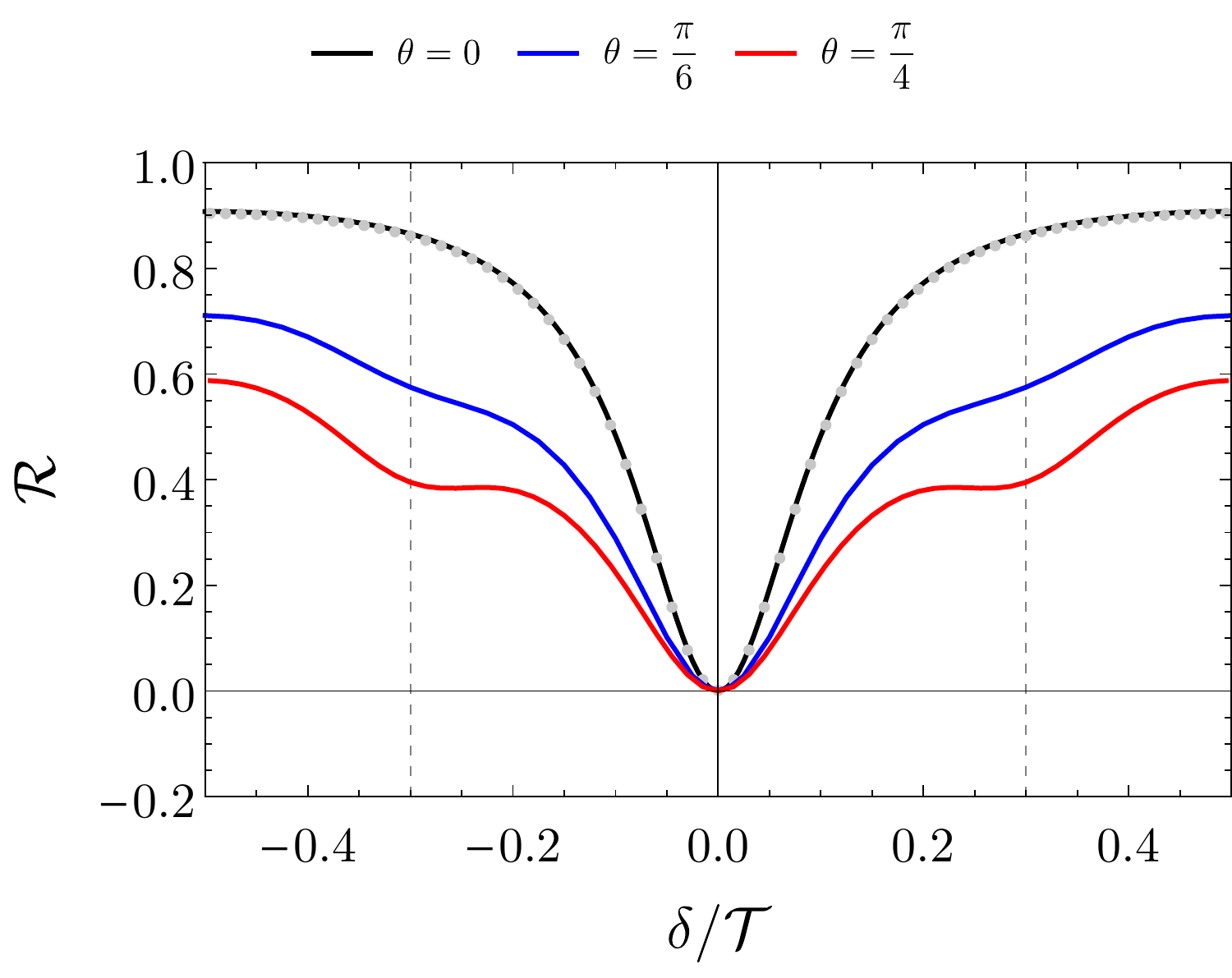}
\caption{\label{fig:fig3a}  Ratio $\mathcal{R}$ in Eq.~\eqref{eq:rapp}, for Lorentzian pulses, as a function of time delay over period ($\delta/\mathcal{T}$) for a symmetric setup. The HOM noise generated by the collision of periodical trains of Lorentzian pulses is shown for different interaction parameters: $\theta=0$ (black curve), $\theta=\pi/6$ (blue curve) and $\theta=\pi/4$ (red curve). Gray dots represent the analytical prediction in Eq.~\eqref{eq:ratio-analytical} for the non-interacting case. Other parameters are: $\tau_0/\mathcal{T}=0.05$, $v_\rho=4\cdot 10^5 m/s$ and $v_\sigma=1.8\cdot 10^5 m/s$, with $L_{\mathrm{A}}=L_{\mathrm{B}}=2 \mu m$. Notice that the positions of side dips occur at $\delta_{\mathrm{sd}}=\pm|\tau_\rho-\tau_\sigma|$ (gray dotted vertical lines).} 
\end{figure}

Due to interactions, as the time delay $\delta$ between the right and the left moving electrons is varied, we find three characteristic features in the noise profile (see Fig.~\ref{fig:fig3a}). At $\delta=0$ a central dip appears while two symmetrical side-dips emerge at positions $\delta_{\mathrm{sd}}=\pm|\tau_\rho-\tau_\sigma|$. The shape of these three dips is Lorentzian reflecting the overall form of the applied voltage pulses, while their width depends on the timescale $\tau_0$. According to this the dips are more pronounced for a smaller ratio $\tau_0/\mathcal{T}$.

This interference pattern is interpreted in terms of the different excitations emerging after the interacting region. Indeed, after the injection, the electronic wave-packet fractionalizes into a slow and a fast mode carrying different charges. According to Eq.~\eqref{eq:new}, the central dip, which corresponds to the situation of simultaneous injection from the two sources, goes exactly to zero because these identical excitations interfere destructively. This is in striking contrast with what has been observed in a HOM experiment at $\nu=2$ where the injection was achieved by means of driven mesoscopic capacitors~\cite{Bocquillon13,Marguerite16}, where the visibility of the central dip is always reduced by interactions~\cite{Wahl14}. 

The destructive interference is also responsible for the side-dip structures appearing when fractionalized excitations with different velocities collide (see Fig.~\ref{fig:fig3a}). For instance, at a delay $\delta_{\mathrm{sd}}=\tau_\sigma-\tau_\rho$ the fast right moving excitation and the slow left moving one reach the QPC at the same time. Furthermore, Fig.~\ref{fig:fig3a} also shows as a reference the behavior of the noise ratio in absence of interactions ($\theta=0$) which always reaches zero (at $\delta=0$) but does not show any side dip because no fractionalization occurs in this case. Our numerical curve (black) perfectly coincide with the theoretical analytical formula (gray dots) derived for the HOM noise ratio $\mathcal{R}$ of colliding Levitons with unitary charge in the absence of interactions ($\theta=0$)~\cite{Dubois13b,Rech17,Ronetti18}
\begin{equation}
    \mathcal{R}_{0}(\delta)=\frac{\sin^2\left(\pi \frac{\delta}{\mathcal{T}}\right)}{\sinh^2\left(2\pi \frac{\tau_0}{\mathcal{T}}  \right)+\sin^2\left(\pi \frac{\delta}{\mathcal{T}}\right)}.
    \label{eq:ratio-analytical}
\end{equation}
We have used this reference result as a check for the validity of our numerical calculations. 

In view of possible future experimental validations of our theoretical analysis, the plots of $\mathcal{R}$ as a function of $\delta/\mathcal{T}$ for different values of $\theta$ have been obtained by fixing the ratio between the pulse width and the period to be $\tau_0/\mathcal{T}=0.05$, compatible with state of the art measurements carried out in narrow constrictions~\cite{Dubois13,Glattli16}, while the time of flight of both slow and fast modes are of the order of $10\div 100$ ps, as interaction lengths are $L\sim \mu m$ and velocities are $v_{\rho/\sigma}\sim 10^{4}\div10^5 m/s$. This makes our prediction observable in nowadays EQO experiments. 


\subsection{\label{sec:asym}Asymmetric setup}
We now examine the HOM noise ratio in Eq.~\eqref{eq:rapp} for an asymmetric configuration where the distances between the injection contacts and the QPC are different ($L_{A}\neq L_{B}$), still assuming the same inter-edge interaction on both arms ($\theta_{A}=\theta_{B}=\theta$). Notice that our general result in Eq.~\eqref{eq:rapp} can be directly used also to investigate the case $\theta_A\neq\theta_B$ even if this condition is more difficult to be controlled experimentally. We did not include this situation in the paper in order to keep the discussion more focused. In any case, we expect different interaction strengths to give a similar qualitative behavior as the presence of different lengths.

Differently from the symmetric case, when $L_A\neq L_B$ the photoassisted coefficients in Eq.~\eqref{eq:rapp} are no longer equal $(\widetilde{p}_{l,A}\neq \widetilde{p}_{l,B})$ because of the different interaction lengths. A new scenario thus emerges in this case as now the right-moving modes and the left-moving ones do not have the same times of flight even if they have the same velocities because of the same interaction strengths. For this reason we denote the times of flight of right-moving modes as $\tau^{A}_{\rho,\sigma}={L_{A}}/{v_{\rho,\sigma}}$ and those of the left-moving ones as $\tau^{B}_{\rho,\sigma}={L_{B}}/{v_{\rho,\sigma}}$. 

\begin{figure}[t]
\includegraphics[width=\columnwidth]{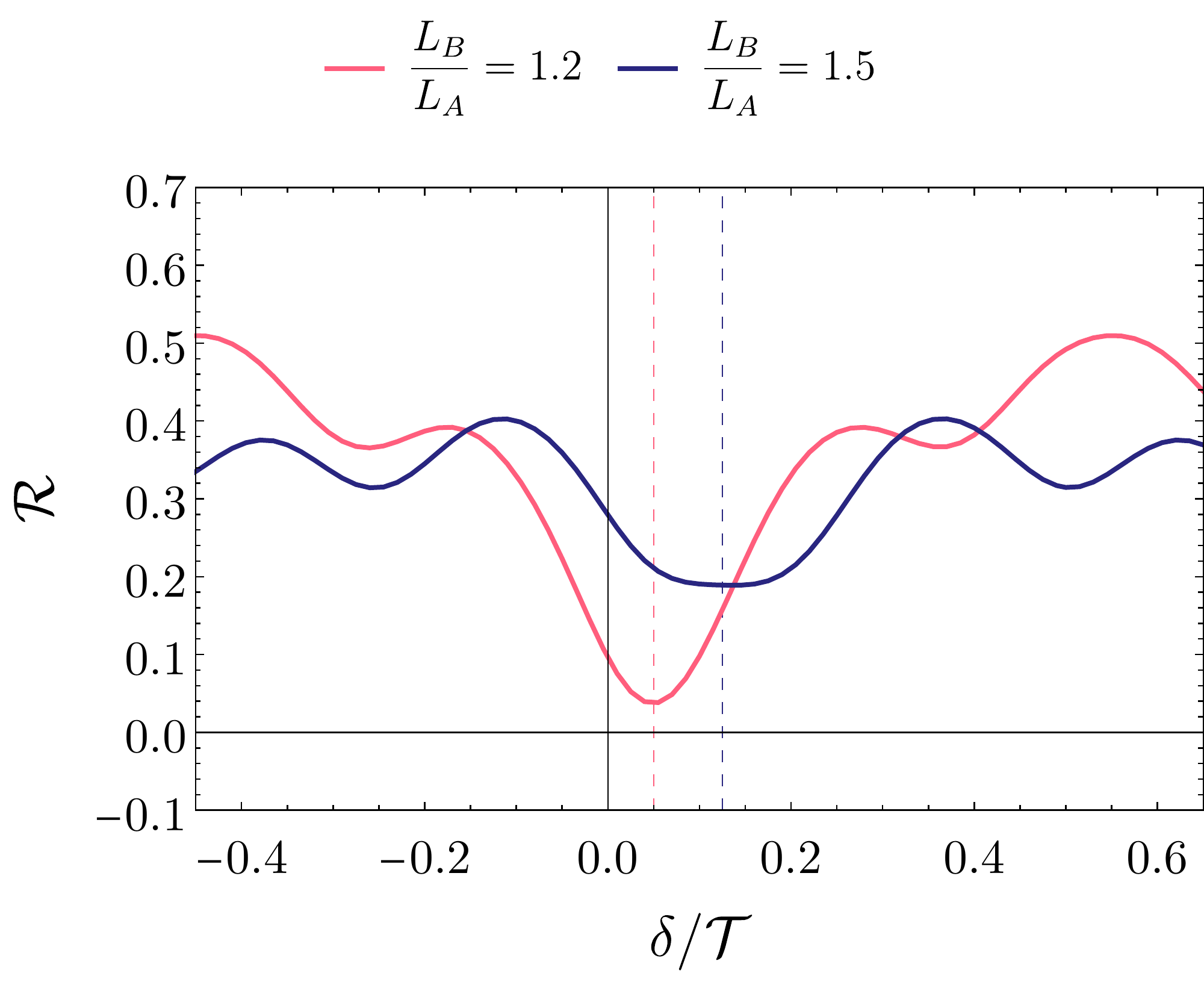}
\caption{\label{fig:fig4} Ratio $\mathcal{R}$ in Eq.~\eqref{eq:rapp}, for Lorentzian pulses, as a function of time delay over period ($\delta/\mathcal{T}$) for an asymmetric setup. In the upper panel, the two curves refer to the strong coupling regime ($\theta=\pi/4$) for two different lengths ratios: $L_{B}/L_{A}=1.2$ (red curve), $L_{B}/L_{A}=1.5$ (blue curve). In the lower one, the curves refer to the non-interacting case ($\theta=0$) for the same lengths ratios as before (red and blue). It is worth noticing that if no interaction occurs the ratio $\mathcal{R}$ goes to zero and that the positions of the minima depend on both the lengths ratios and the velocity propagation of the free fermions along the channel of the injection (here assumed to be $v_1=v_\rho$). Other parameters are: $\tau_0/\mathcal{T}=0.05$, $v_\rho=4\cdot 10^5 m/s$ and $v_\sigma=1.8\cdot 10^5 m/s$, with $L_{\mathrm{A}}=2 \mu m$. Notice that the positions of central dip are indicated respectively by the red and blue dashed vertical lines.}
\end{figure}

As before, we consider the noise ratio $\mathcal{R}$ as a function of the time delay $\delta$ focusing on the strong coupling regime ($\theta=\pi/4$) and considering different values of the length ratio $L_{B}/L_{A}$. From Fig.~\ref{fig:fig4} (upper panel) one can outline that the three dips described before are still present, but now the overall profiles are very different with respect to the symmetric case. Indeed, here the central dip does not reach anymore zero (loss of visibility) and its position is shifted with respect to the origin by a time delay
\begin{equation}
\delta_{\mathrm{cd}}=\frac{\tau^{B}_\sigma+\tau^{B}_\rho-\tau^{A}_\sigma-\tau^{A}_\rho}{2}
\label{eq:delta-cd}
\end{equation}
which increases proportionally to the length ratio $L_{B}/L_{A}$. This means that the total suppression of HOM noise is not achieved because the different interaction lengths result in different times of flight ($\tau_{\rho,\sigma}^A\neq\tau_{\rho,\sigma}^B$) in such a way that the charge and neutral parts of the incoming signals do not reach the QPC at the same time. The distances of the side dips from the central one satisfy
\begin{equation}
|\delta_{\mathrm{cd}}-\delta_{\mathrm{sd}}|=\frac{\tau^{B}_\sigma-\tau^{B}_\rho+\tau^{A}_\sigma-\tau^{A}_\rho}{2}
\label{eq:sidedip}
\end{equation} 
clearly showing the effect of the asymmetric lengths of interacting regions on the noise ratio $\mathcal{R}$.\\
Notice that in absence of interactions (lower panel of Fig.~\ref{fig:fig4}) we recover the same behavior of the symmetric case in Fig.~\ref{fig:fig3a} up to a simple shift in the delay direction. This is a direct consequence of the lack of fractionalization of the incoming excitations.

An additional comment on Eq.~\eqref{eq:delta-cd} is worthwhile. In the symmetric setup, the central dip corresponds to the situation where the two charged (or dipolar) modes incoming from the two sources arrive simultaneously at the QPC. In the asymmetric case, at a delay $\delta_1=\tau_\rho^B-\tau_\rho^A$ ($\delta_2=\tau_\sigma^B-\tau_\sigma^A$) the charged (dipolar) modes reach the QPC at the same time, but the dipolar (charged) ones do not. As a result, instead of a single central dip as appearing in Fig.~\ref{fig:fig4}, two distinct dips located at $\delta_1$ and $\delta_2$ should be expected (for additional details see Appendix~\ref{app3}). However, for realistic values for $\tau_0/\mathcal{T}$, these two dips are not resolved (because the wavepackets are not narrow enough) and merge into a broader one, located at an average delay $\delta_{\mathrm{cd}}=(\delta_1+\delta_2)/2$.

\begin{figure}[t]
\includegraphics[width=\columnwidth]{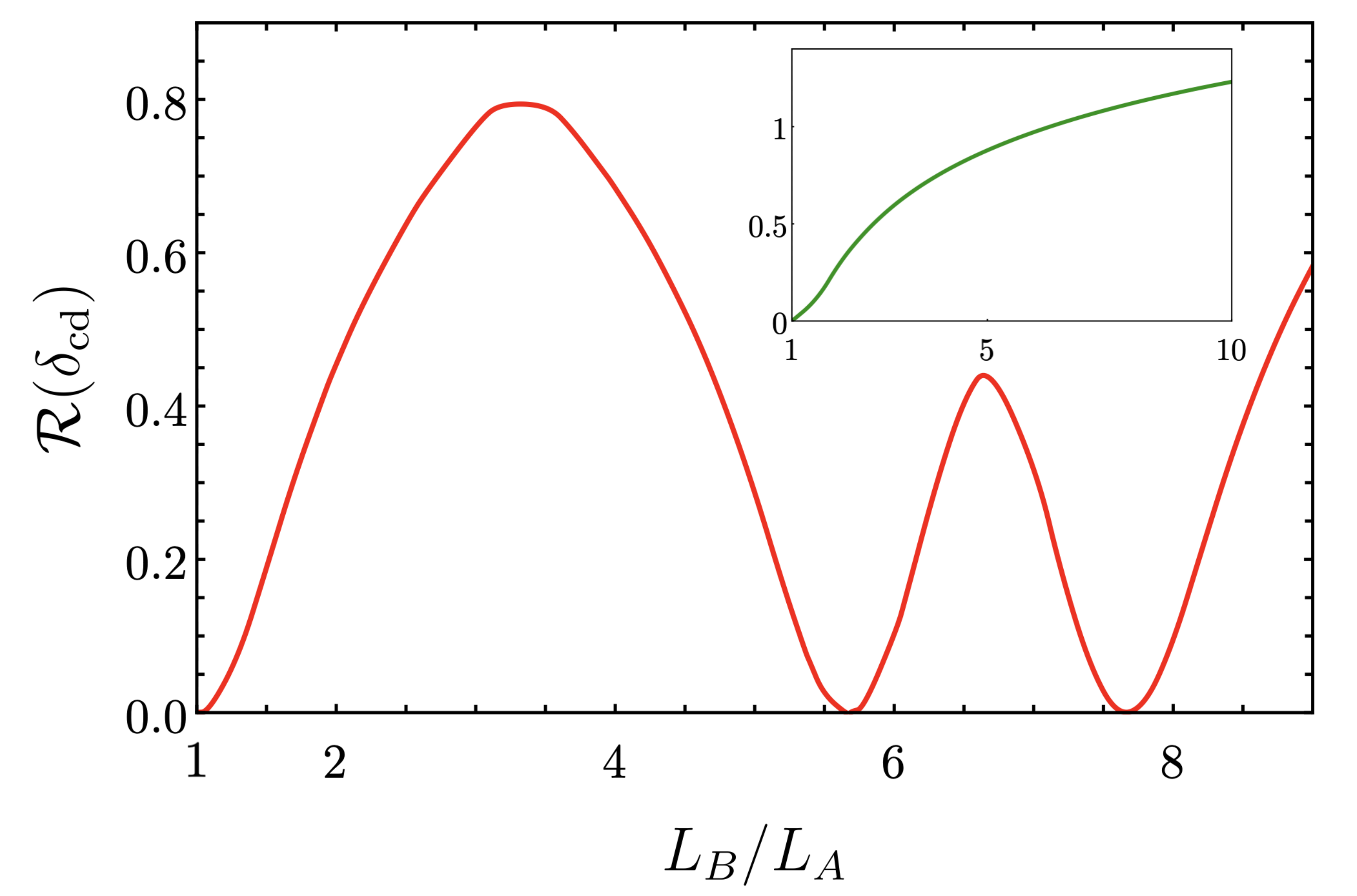}
\caption{\label{fig:fig5} Behavior of $\mathcal{R}(\delta_\mathrm{cd})$ as a function of the lengths ratio $L_\mathrm{B}/L_\mathrm{A}$. The red curve is obtained for a periodic Lorentzian pulse ($\tau_0/\mathcal{T}=0.05$) while the green inset shows the case of a single pulse. Other parameters are: $\theta=\pi/4$, $L_\mathrm{A}=2\mu m$, $v_\rho=1.5\cdot 10^5 m/s$ and $v_\sigma=2\cdot 10^4 m/s$. Notice that here we have chosen propagation velocities different with respect to the other Figures with the only aim of magnifying the features discussed in the main text.}
\end{figure}

In Section \ref{sec:symmetric} we have shown that the HOM noise goes exactly to zero when we are in a symmetric situation and the excitations are injected simultaneously in the QH edge channels. This time one may wonder whether the signal periodicity affects the visibility of the central dip ($\mathcal{R}(\delta_{\mathrm{cd}})$) in an asymmetric setup when the lengths ratio is varied. In Fig.~\ref{fig:fig5} we show the behavior of the minimum of the HOM ratio ($\mathcal{R}(\delta_{\mathrm{cd}})$) as a function of $L_{B}/L_{A}$ in the presence of a periodical Lorentzian source (main plot) and compare it to the single Lorentzian pulse case (inset). Also in this case we focus on the strong coupling $\theta=\pi/4$ regime. The biggest difference between the two situations lies in the fact that for the periodic drive the red curve goes to zero three times in the considered range of the ratio $L_{B}/L_{A}$, including the starting point (where $L_{B}/L_{A}=1$), while for the single pulse no other zero occurs apart the one corresponding to equal lengths. Therefore, the occurrence of additional zeros is a remarkable consequence of the periodicity of the drive and can be used to extract information about the interaction parameter $\theta$.

In order to better understand the behavior in Fig.\ \ref{fig:fig5} we must start from Eq.~\eqref{eq:new}, for a generic case with $\delta\neq 0$. Therein, the phases $\varphi_A$ and $\varphi_B$ must be equal in order to have a perfect superposition of colliding excitations and a consequent maximal visibility of the HOM central dip. 
The expression giving the lengths ratios at which the central dip is maximally visible in the case of a periodical injection is
\begin{equation}
    \frac{L_{\mathrm{B}}}{L_{\mathrm{A}}}=\frac{2(k-k')\mathcal{T}}{\tau_\sigma^A-\tau_\rho^A}+1
    \label{eq:long}
\end{equation}
with $k,k'\in \mathbb{N}$ and $k>k'$ (see Appendix \ref{app2} for more details). The previous relation describes the zero located at $L_{\mathrm{B}}/L_{\mathrm{A}}\approx 7.6$ in Fig.~\ref{fig:fig5}, for $k-k'=1$. We also point out that the presence of a second zero, located at $L_{\mathrm{B}}/L_{\mathrm{A}}\approx 5.6$, is a direct consequence of the maximal coupling $\theta=\pi/4$. Indeed, as shown in Appendix \ref{app2}, in this condition additional zeros appear for a length ratio
\begin{equation}
    \frac{L_{\mathrm{B}}}{L_{\mathrm{A}}}=\frac{2(k-k')\mathcal{T}}{\tau_\sigma^A-\tau_\rho^A}-1\,.
    \label{eq:long1}
\end{equation}
However, as soon as the coupling departs from the maximal value ($\theta<\pi/4$) the second zero is lifted and turns into a local minimum (see Appendix~\ref{app2} for more details). This is a signature of the different weight of charge and dipole contributions to the fractionalized wave-packet and can be use to extract information about the mismatch in the time of flight and consequently about the inter-edge coupling $\theta$.

The possibility for the HOM central dip to reach zero at different values of the lengths ratio is a direct consequence of the periodicity of the applied signal. In terms of electronic density we can think of what is happening as follows: one Leviton is injected for every period, it crosses the interacting region where it fractionalizes into two modes with different velocities. If the interacting region has the proper length, the fast mode of a given period will reach the slow mode of the previous one. By properly calibrating the ratio between the lengths into the two arms it is possible to achieve a situation where the colliding objects, the fast and slow modes coming from both arms, at the QPC are identical leading to a vanishing HOM noise.


\section{Measuring interactions}\label{sec:meas}
Until now, we have considered a setup where the injection takes place on the inner channels of QH bar only. A more general analysis consists in considering a case where the excitations are also injected in the outer channels. This configuration can be achieved for example by further exploiting an open quantum dot coupled to the outer channels \cite{Bocquillon13}. As we will demonstrate, in this case a collisional HOM experiment allows to extract information on the interaction strength between the edge channels as long as it can be assumed as short-range. In Section \ref{sec:asym} we have shown how the dependence of the visibility of the central dip as a function of the lengths ratio can be used to indirectly estimate interactions. Here, we consider a more direct way to measure the interaction intensity encoded in the parameter $\theta$.

Let us start by considering two different input voltages $V^{A/B}_{1/2}$ at the entrance of the interaction region, where $1$ stands for the inner channels and $2$ for the outer ones. Without loss of generality we consider the two drives to be proportional, namely $V^{A/B}_{2,\mathrm{in}}=\alpha V^{A/B}_{1,\mathrm{in}}$. In what follows we only consider a symmetric configuration for the interferometer even if similar results can be obtained for an asymmetric case. This implies that Eq.~\eqref{eq:voltage} can be written as
\begin{equation}
 \begin{split}
      V^{A/B}_{1,\mathrm{out}}(t)=&\cos^2{\theta}\,V^{A/B}_{1,\mathrm{in}}(t-\tau_\rho)+\sin^2{\theta}\,V^{A/B}_{1,\mathrm{in}}(t-\tau_\sigma)\\ +&\alpha\sin{\theta}\cos{\theta}[V^{A/B}_{1,\mathrm{in}}(t-\tau_\rho)-V^{A/B}_{1,\mathrm{in}}(t-\tau_\sigma)]\\
      V^{A/B}_{2,\mathrm{out}}(t)=&\sin{\theta}\cos{\theta}[V^{A/B}_{1,\mathrm{in}}(t-\tau_\rho)-V^{A/B}_{1,\mathrm{in}}(t-\tau_\sigma)]\\ +&\alpha\sin^2{\theta}\,V^{A/B}_{1,\mathrm{in}}(t-\tau_\rho)+\alpha\cos^2{\theta}\,V^{A/B}_{1,\mathrm{in}}(t-\tau_\sigma).
 \end{split}
 \label{eq:alpa}
\end{equation}
From the above equation, we can identify two relevant situations involving two different values of the proportionality parameter: $\alpha=\tan\theta$ and $\alpha=-\cot\theta$. For these two values $(V^{A/B}_{1,\mathrm{in}},\alpha V^{A/B}_{1,\mathrm{in}})^T$ is an eigenvector of the scattering matrix $\mathcal{S}$ in Eq.~\eqref{eq:smatrix}. In the time domain, this results in
\begin{equation}
    \begin{pmatrix}
    V^{A/B}_{1,\mathrm{out}}(t)\\
    V^{A/B}_{2,\mathrm{out}}(t)
    \end{pmatrix}=
    \begin{pmatrix}
    V^{A/B}_{1,\mathrm{in}}(t-\tau_\rho)\\
    V^{A/B}_{2,\mathrm{in}}(t-\tau_\rho)
    \end{pmatrix}\quad\text{for }\alpha=\tan\theta
\end{equation}
and
\begin{equation}
    \begin{pmatrix}
    V^{A/B}_{1,\mathrm{out}}(t)\\
    V^{A/B}_{2,\mathrm{out}}(t)
    \end{pmatrix}=
    \begin{pmatrix}
    V^{A/B}_{1,\mathrm{in}}(t-\tau_\sigma)\\
    V^{A/B}_{2,\mathrm{in}}(t-\tau_\sigma)
    \end{pmatrix}\quad\text{for }\alpha=-\cot\theta\,.
\end{equation}

Therefore, for these values of $\alpha$, the input voltages are not mixed by interactions and are transferred unaffected to the output of the interacting region. This feature is quite surprising because it means that, by properly tuning $\alpha$, one can inject two input excitations which effectively propagate freely along the edge channels without undergoing any fractionalization process, despite the presence of an interacting region in the system. Therefore, it is possible to regard $\alpha$ as a tunable parameter with which one can switch off interaction effects on the HOM noise ratio $\mathcal{R}$. As a possible experimental way to implement such kind of voltage configuration one can apply the same voltage $V^{A/B}_{1, \mathrm{in}}(t)$ to both channels, further adding a voltage $(\alpha-1)V^{A/B}_{1, \mathrm{in}}(t)$ properly synchronized with the first one only to channel $2$ by means of a quantum dot~\cite{Bocquillon13}.

\begin{figure}[t]
\includegraphics[width=\columnwidth]{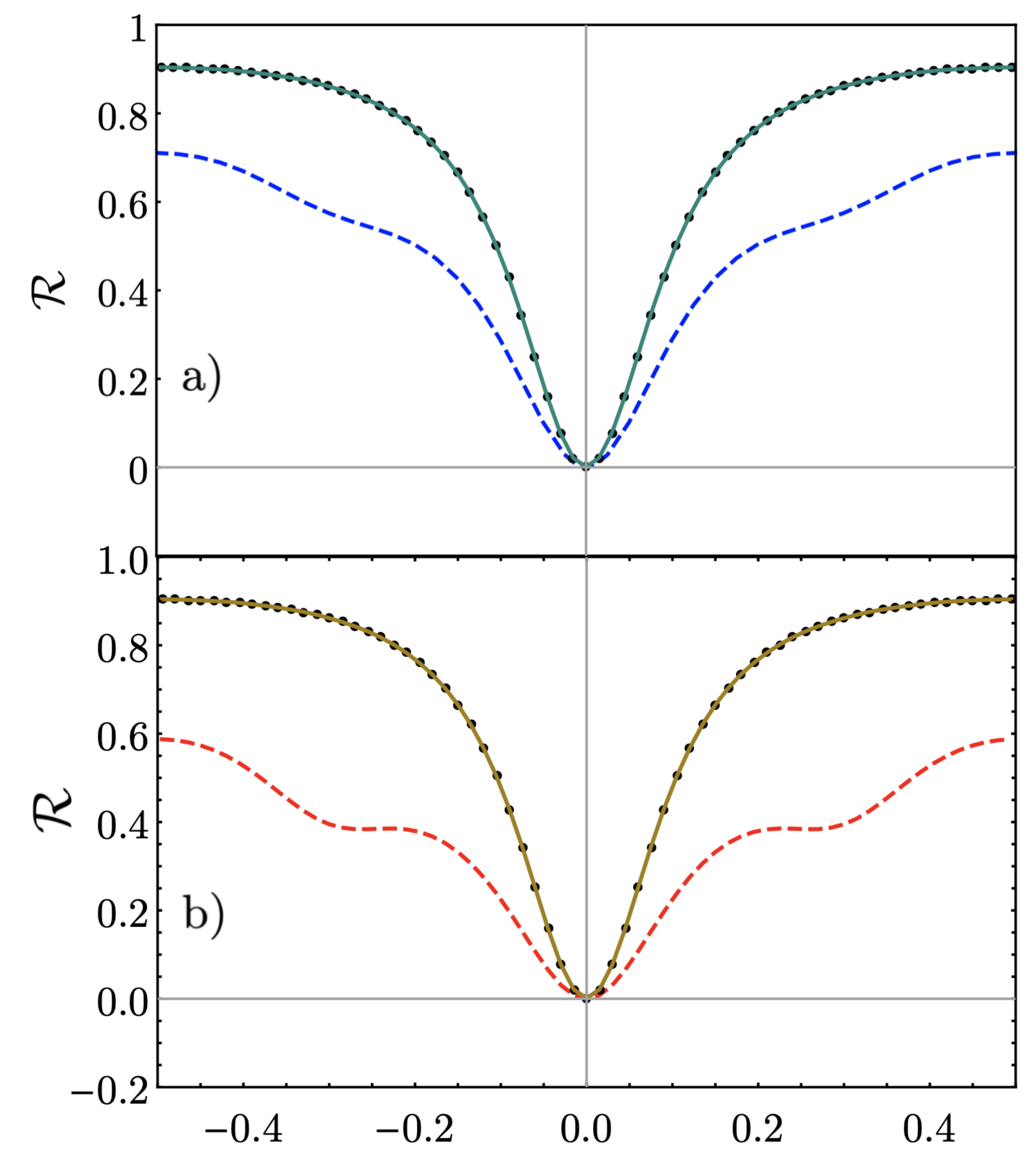}
\caption{\label{fig:fig6} Ratio $\mathcal{R}$ as a function of time delay over period ($\delta/\mathcal{T}$). The full lines describe HOM collisions at finite values of $\alpha$. In a) we set $\theta=\frac{\pi}{6}$ and $\alpha=\tan\left(\frac{\pi}{6}\right)\approx 0.58$ while in b)  $\theta=\frac{\pi}{4}$ and $\alpha=\tan\left(\frac{\pi}{4}\right)=1$. The dashed lines describe the corresponding curves for an injection only in inner channels ($\alpha=0$) for both a) $\theta=\frac{\pi}{6}$ and b) $\theta=\frac{\pi}{4}$). The black dots show the ratio $\mathcal{R}$ in the absence of interactions ($\theta=0$). Other parameters are: $\tau_0/\mathcal{T}=0.05$, $v_\rho=4\cdot 10^5 m/s$ and $v_\sigma=1.8\cdot 10^5 m/s$, with $L_{\mathrm{A}}=L_{\mathrm{B}}=2 \mu m$.}
\end{figure}

In order to illustrate this effect we compare in Fig.\ \ref{fig:fig6} the case where the injection only occurs in the inner channels ($\alpha=0$) with the situation when both inner and outer channels are driven ($\alpha\neq0$). The former scenario is represented by the dashed curves, showing the side dips structure already discussed in Section \ref{sec:symmetric}. The latter case is represented by full lines and clearly shows that, for the particular value $\alpha=\tan\theta$, the side dips disappear and one perfectly recovers the same behavior as in the absence of interactions, described by Eq.\ \eqref{eq:ratio} and shown by black dots in Fig.\ \ref{fig:fig6}. 

As a final remark we mention that, for $\alpha>\tan\theta$ (not shown), the side dips become side peaks as a consequence of the fact that excitations with opposite charge reach the QPC~\cite{Jonckheere12,Wahl14}. 

According to the above considerations it is clear that the study of the evolution of the side dips as a function of $\alpha$ can be used as a way to estimate the value of the inter-edge interaction. Indeed, by tuning $\alpha$ in such a way to eliminate the side dips in the HOM signal knowing that this occurs precisely at $\alpha=\tan\theta$, the mixing angle can be obtained from this relation.

\section{\label{sec:concl} Conclusions}
We have theoretically investigated a Hong-Ou-Mandel experiment where periodic time-dependent voltage drives are injected with a tunable delay into Quantum Hall edge channels at filling factor $\nu=2$ and collide at a Quantum Point Contact.
In particular, we have focused on Lorentzian voltage pulses carrying unitary charge, usually dubbed Levitons. As a consequence of the screened Coulomb interactions between the edge channels, the injected electrons fractionalize leading to an interesting phenomenology. Indeed, the noise measured just outside the Quantum Point Contact is characterized by the emergence of side dips as a function of the delay in the injection. Moreover, differently from what happens in the case of injection using driven mesoscopic capacitors, the visibility of the central dip remains maximal independently of the interaction for a symmetric device \textit{regardless} of the form of the voltage used for the injection. This fact is a signature of the robustness of voltage drives with respect to interaction effects.
Our results are even more interesting in the case of Lorentzian voltage pulses because of their relevance as on-demand single electron sources. Only by inducing an asymmetry in the device, for example by considering different distances between the injectors and the Quantum Point Contact, the visibility can be reduced. In addition, from the peculiar dependence of the visibility on the ratio between these distances, it is possible to extract information about the intensity of the interaction along the edge. Along this direction, we have also proposed a more direct measurement of the interaction based on the application of different voltages on the two edge channels along each arm of the interferometer. In this case, by properly tuning the ratio between these voltages, it is possible to prevent the fractionalization with a consequent disapperance of the side dip in the HOM noise profile. Therefore, the study of the evolution of the side dips in this configuration can provide a direct measurement of the strength of inter-edge interaction.   

\acknowledgements
The authors would like to thank T. Martin, T. Jonckheere, J. Rech, F. Ronetti and J. Splettstoesser for useful discussions. M.A.~acknowledges support from the European Union’s H2020 research and innovation programme under grant agreement No 862683.
\appendix

\section{Photoassisted amplitudes and HOM noise ratio}
\label{app1}
In this Appendix we evaluate the photoassisted amplitudes $\widetilde{p}_{l,A/B}(q)$ and $\mathcal{P}_l(q;\delta)$ in terms of amplitudes $p_l(q)$ and we show how to obtain Eq.~\eqref{eq:rapp}. The coefficients we want to determine, for a periodic voltage pulse source $V(\tau)=V(\tau+\mathcal{T})$, are defined by the Fourier series $(j=A,B)$
\begin{equation}
    e^{-ie\int_0^t V_{1,\mathrm{out}}^{j}(\tau)d\tau}=e^{-iq\Omega t}\sum_l\widetilde{p}_{l,j}(q)e^{-il\Omega t}
    \label{eq:def-ptilde}
\end{equation}
where $\Omega=2\pi/\mathcal{T}$ and
\begin{equation}
    e^{-ie\int_0^t \left[V_{1,\mathrm{out}}^{A}(\tau)-V_{1,\mathrm{out}}^{B}(\tau)\right]d\tau}=\sum_l\mathcal{P}_{l}(q;\delta)e^{-il\Omega t}\,.
    \label{eq:def-pcheck}
\end{equation}
The explicit expressions of these coefficients in terms of $p_l$ are then obtained by inverting the previous relations. Let us start with $\tilde{p}_{l,j}(q)$. From Eq.~\eqref{eq:def-ptilde} we have
\begin{equation}
    \widetilde{p}_{l,j}(q)=\int_0^\mathcal{T}\frac{dt}{\mathcal{T}}e^{i(l+q)\Omega t}e^{-ie\int_0^t V_{1,\mathrm{out}}^{j}(\tau)d\tau}.
    \label{eq:p-tilde-inverse}
\end{equation}
Next, with the help of Eqs.~\eqref{eq:relation-va-vb} and~\eqref{eq:alpa}, the voltages $V_{1,\mathrm{out}}^{j}(t)$ are expressed in terms of the source drive $V(t)$ given in Eq.~\eqref{eq:lor}. In doing that, four different phase factors involving $V(t)$ are obtained, each of which containing a time shift and being differently weighted due to interactions. It is then possible to repeatedly use Eq.~\eqref{eq:four} to express these factors as Fourier series involving the photoassisted coefficients $p_l$. Finally, after performing the time integration in Eq.~\eqref{eq:p-tilde-inverse} we obtain the result (neglecting unimportant phases)
\begin{equation}
\begin{split}
    \widetilde{p}_{l,j}(q)&=\sum_{nrs}p_{l-n-r+s}(q_1)p_n(q_2)p_r(q_3)p_s^*(q_3)\\
    &\quad\times e^{i\Omega\tau_\rho^j(l-n+s)}e^{i\Omega\tau_\sigma^j(n-s)},
    \end{split}
    \label{eq:coeff2}
\end{equation}
where $q_1=\cos^2\theta$, $q_2=\sin^2\theta$ and $q_3=\alpha\sin\theta\cos\theta$. Thus, the coefficients $\widetilde{p}_{l,j}(q)$ are completely specified once the expression of $p_l(q)$ is known. For a Lorentzian drive, it is given by~\cite{Dubois13,Grenier13,Rech17,Ferraro18}
\begin{equation}
p_l(q)=q\sum_{s=0}^{+\infty}\frac{(-1)^s\Gamma(q+l+s)e^{-2\pi\tau_0(2s+l)/\mathcal{T}}}{\Gamma(q+1-s)\Gamma(1+s)\Gamma(1+l+s)}.
\label{eq:coefflor}
\end{equation}
Once $\widetilde{p}_{l,j}$ are known, it is easy to obtain the photoassisted coefficients $\mathcal{P}_{l}(q;\delta)$ that take into account the time delay $\delta$ between the two sources. Indeed, by inverting Eq.~\eqref{eq:def-pcheck} and using Eq.~\eqref{eq:def-ptilde} to express the phase factors involving the voltages $V_{1,\mathrm{out}}^j(t)$, we readily arrive at the expression
\begin{equation}
    \mathcal{P}_l(q;\delta)=\sum_{m}\widetilde{p}_{l+m,A}(q)\widetilde{p}^*_{m,B}(q)\, e^{i m\Omega\delta},
    \label{eq:coeffp}
\end{equation}
where, again, unimportant phases have been neglected.

Now we have all the ingredients to evaluate the HOM ratio defined in Eq.~\eqref{eq:ratio}. Recall that $S_\mathrm{HOM}$ is obtained by evaluating Eq.~\eqref{eq:newnoise} and the contributions $S_{\mathrm{HBT},j}$ are particular cases when one of the two sources is switched off. The first step is to express the phases $\varphi_j(t,t')$ by relying on the photoassisted coefficients we have determined in this Appendix. For instance,
\begin{equation}
\begin{split}
   e^{-i\varphi_A(\bar{t}+\frac{\tau}{2},\bar{t}-\frac{\tau}{2})} &=\bigg(\sum_l \widetilde{p}_{l,A}(q) \, e^{-i l\Omega(\bar{t}+\frac{\tau}{2})} e^{-i q\Omega(\bar{t}+\frac{\tau}{2})}\bigg)\\
   &\times \bigg(\sum_{l'} \widetilde{p}_{l',A}^*(q) \, e^{i l'\Omega(\bar{t}-\frac{\tau}{2})} e^{i q\Omega(\bar{t}-\frac{\tau}{2})}\bigg) \\
   &=\sum_{ll'}\widetilde{p}_{l,A}(q)\widetilde{p}_{l',A}^*(q)\,e^{i\Omega\bar{t}(l'-l)}e^{-i\Omega\frac{\tau}{2}(l+l'+2q)}
   \end{split}
\end{equation}
and similarly for $e^{-i\varphi_B}$, where the time delay $\delta$ has to be taken into account.
This expression is then used into Eq.~\eqref{eq:new} to obtain the function $\Delta Q(t+\tau/2,t-\tau/2)$. Finally, the two time integrations in Eq.~\eqref{eq:new} can be performed yielding (in the limit of zero temperature)
\begin{equation}
    S_{\mathrm{HOM}}=-(ev_F)^2 RT\bigg(\pi\sum_{l}|\mathcal{P}_l(q;\delta)|^{2}|\Omega l|\bigg)
    \label{eq:ap}
\end{equation}
for the general HOM case and
\begin{equation}
    S_{\mathrm{HBT},j}=-(ev_F)^2 RT\bigg(\pi\sum_{l}|\widetilde{p}_{l,j}(q)|^{2} |\Omega(l+q)| \bigg)
\end{equation}
for the HBT contributions. From these expressions, the noise ratio $\mathcal{R}$ in Eq.~\eqref{eq:rapp} follows straightforwardly.
All the above (infinite) sums are convergent and their value has been obtained numerically by summing over a finite number of coefficients until the desired precision is obtained. 
\section{Central dip resolution in the asymmetric length case}
\label{app3}

\begin{figure}[t]
\includegraphics[width=\columnwidth]{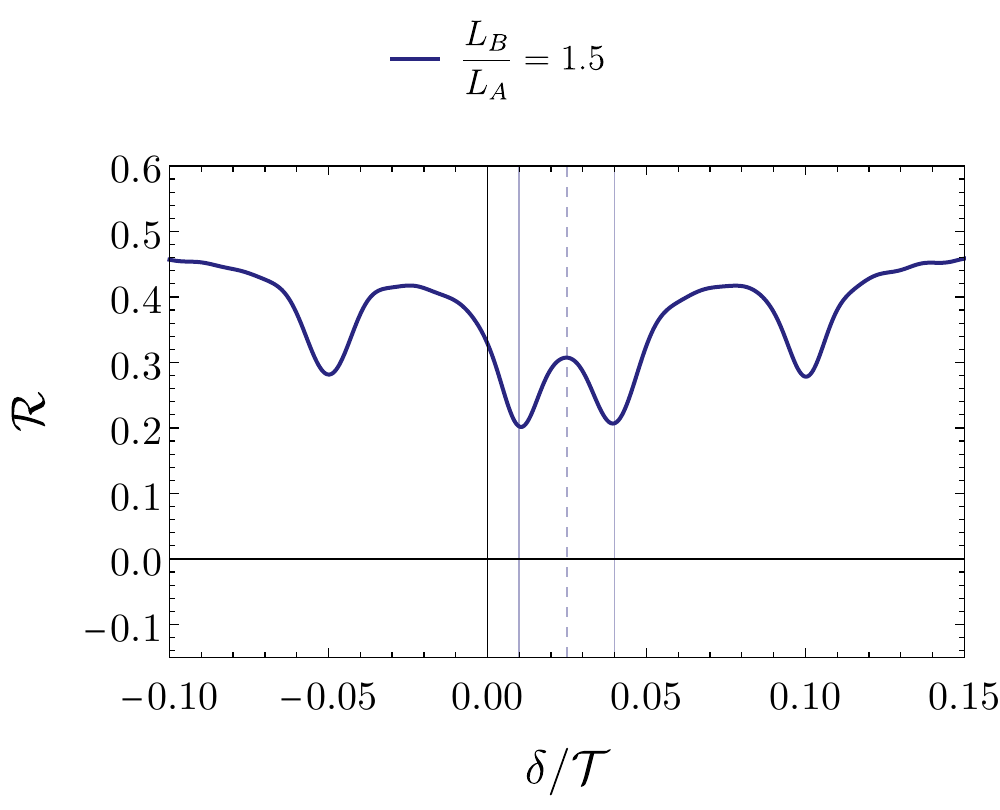}
\caption{\label{fig:app3} Ratio $\mathcal{R}$ in Eq.~\eqref{eq:delta-cd}, for Lorentzian pulses, as a function of time delay over period ($\delta/\mathcal{T}$) for an asymmetric setup. The lengths ratio of the two arms is $L_B/L_B=1.5$, the channel coupling strength is maximal ($\theta=\pi/4$) and one fix $\mathcal{T}=100\tau_0$. Vertical solid lines correspond respectively to $\delta_1=\tau_{\rho}^B-\tau_{\rho}^A$ and $\delta_2=\tau_{\sigma}^B-\tau_{\sigma}^A$ while the dotted vertical one to $\delta_{cd}$ (in Eq.~\eqref{eq:delta-cd}). Other parameters are: $\tau_0/\mathcal{T}=0.05$, $v_\rho=4\cdot 10^5 m/s$ and $v_\sigma=1.8\cdot 10^5 m/s$, with $L_{\mathrm{A}}=2 \mu m$. } 
\end{figure}

In this Appendix we want to comment the resolution of the central dip in the asymmetric case $\theta_A=\theta_B$ and $L_A\neq L_B$. \\
In principle when the interferometer is asymmetric, one should expect four dips, whose positions would be indeed determined by two delay times only. At $\delta_1=\tau_{\rho}^B-\tau_{\rho}^A$, the two charged modes arrive at the QPC simultaneously, but the dipolar ones do not. Similarly, at $\delta_2=\tau_{\sigma}^B-\tau_{\sigma}^A$, the two dipolar modes arrive at the QPC at the same time while the charged ones do not. However, the proper visualization of those dips requires enough resolution, namely the wave-packets have to be narrow enough. This feature is simply not resolved in Fig.~\ref{fig:fig4} (upper panel) where we have chosen a value for $\tau_0$ (width of the Lorentzian voltage) and $\mathcal{T}$ (period of the source) in the typical range accessible for the experiments. As a result, the two expected dips at $\delta_1$ and $\delta_2$ merge into a broader one, located at $\delta_{cd}$ (Eq.~\eqref{eq:delta-cd}), which is the average delay between the previous two: $\delta_{cd}=(\delta_1+\delta_2)/2$. In Fig.~\ref{fig:app3} we show how the ratio $\mathcal{R}$ should be if we consider unrealistically narrow pulses (for a period $\mathcal{T}=100\tau_0$). Here the two dips are well resolved respectively at positions $\delta_1$ and $\delta_2$ symmetrically with respect to $\delta_{cd}$, bringing the number of observed dips in the HOM ratio from three (as in Fig.~\ref{fig:fig4}) to four.
Concerning the sideband dips, they are located at $\delta=\tau_{\sigma}^B-\tau_{\rho}^A$ (coincidence between the dipolar mode incoming from B and the charged one incoming from A) and $\delta=\tau_{\rho}^B-\tau_{\sigma}^A$ (coincidence between the charged mode incoming from B and the dipolar one incoming from A). Therefore their positions are determined by two times only. In Eq.~\eqref{eq:sidedip} we have expressed the location of these side dips relative to the broad central one.\\
It is worth noticing that in the symmetric situation when $L_A=L_B$ we have $\delta_1=\delta_2=\delta_{cd}=0$ as expected.

\section{Central dip visibility in the asymmetric length case}
\label{app2}
\begin{figure}[t]
\includegraphics[width=\columnwidth]{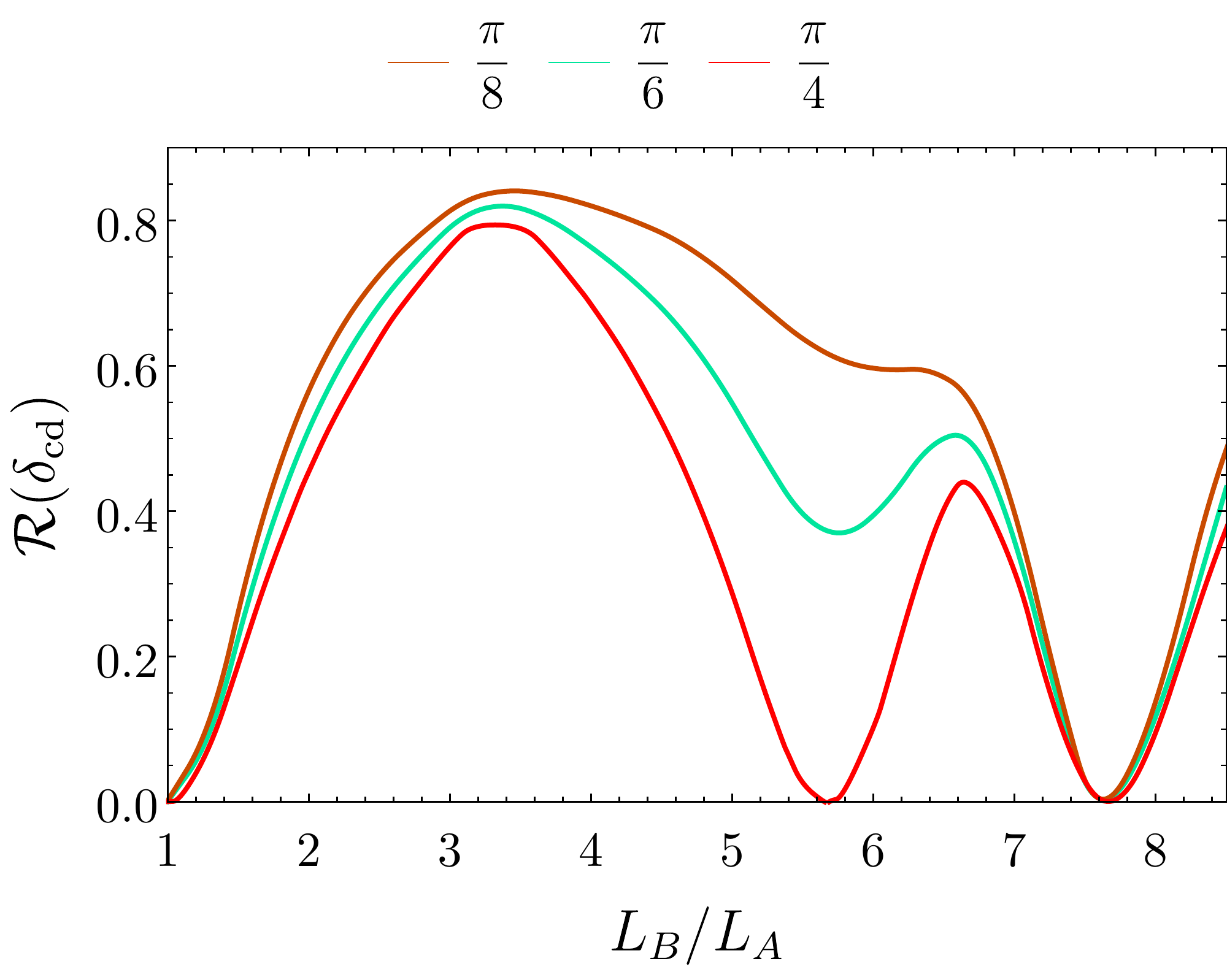}
\caption{\label{fig:fig7} Behavior of $\mathcal{R}(\delta_\mathrm{cd})$ with respect to lengths ratio $L_\mathrm{B}/L_\mathrm{A}$ for three different fixed value of the interaction angle: $\theta=\frac{\pi}{4}$ (red curve), $\theta=\frac{\pi}{6}$ (cyan curve) and $\frac{\pi}{8}$ (brown curve). Other parameters are: $\tau_0/\mathcal{T}=0.05$, $L_A=2 \mu m$ $v_\rho=1.5\cdot 10^5 m/s$ and $v_\sigma=2\cdot 10^4 m/s$.}
\end{figure}
The result in Eq.~\eqref{eq:long}, relating the lengths ratio to the periodicity of the signal used for explanation of the minimum value for $\mathcal{R}(\delta_{\mathrm{cd}})$ in Fig.~\ref{fig:fig5}, is obtained starting from the equality in Eq.~\eqref{eq:relation-va-vb}, that is still true also for the phases $\varphi_j(t)$. The phases are (considering the injection only in one channel)
\begin{equation}
    \begin{split}
        \varphi_{A}(t)&=\cos^2\theta\,\varphi(t-\tau^{A}_\rho)+\sin^2\theta\, \varphi(t-\tau^{A}_\sigma)\\
         \varphi_{B}(t+\delta)&=\cos^2\theta \varphi(t-\tau^{B}_\rho+\delta)+\sin^2\theta \varphi(t-\tau^{B}_\sigma+\delta)
    \end{split}
    \label{eq:mi}
\end{equation}
where $\varphi(t)=\sum_{k\in\mathbb{Z}}\arctan\left( \frac{t-k\mathcal{T}}{\tau_0}\right)$ for Lorentzian periodic pulses. In order to solve $\varphi_\mathrm{A}(t)=\varphi_\mathrm{B}(t+\delta)$ we have to specify the interaction angle. Firstly considering $0<\theta<\frac{\pi}{4}$ (Fig.~\ref{fig:fig7}) we know that $\sin\theta\neq\cos\theta$ and this means that Eq.~\eqref{eq:relation-va-vb} is satisfied when
\begin{equation}
    \varphi(t-\tau^{A}_{\rho/\sigma})=\varphi(t-\tau^{B}_{\rho/\sigma}+\delta)
\end{equation}
\begin{figure}[t]
\includegraphics[width=\columnwidth]{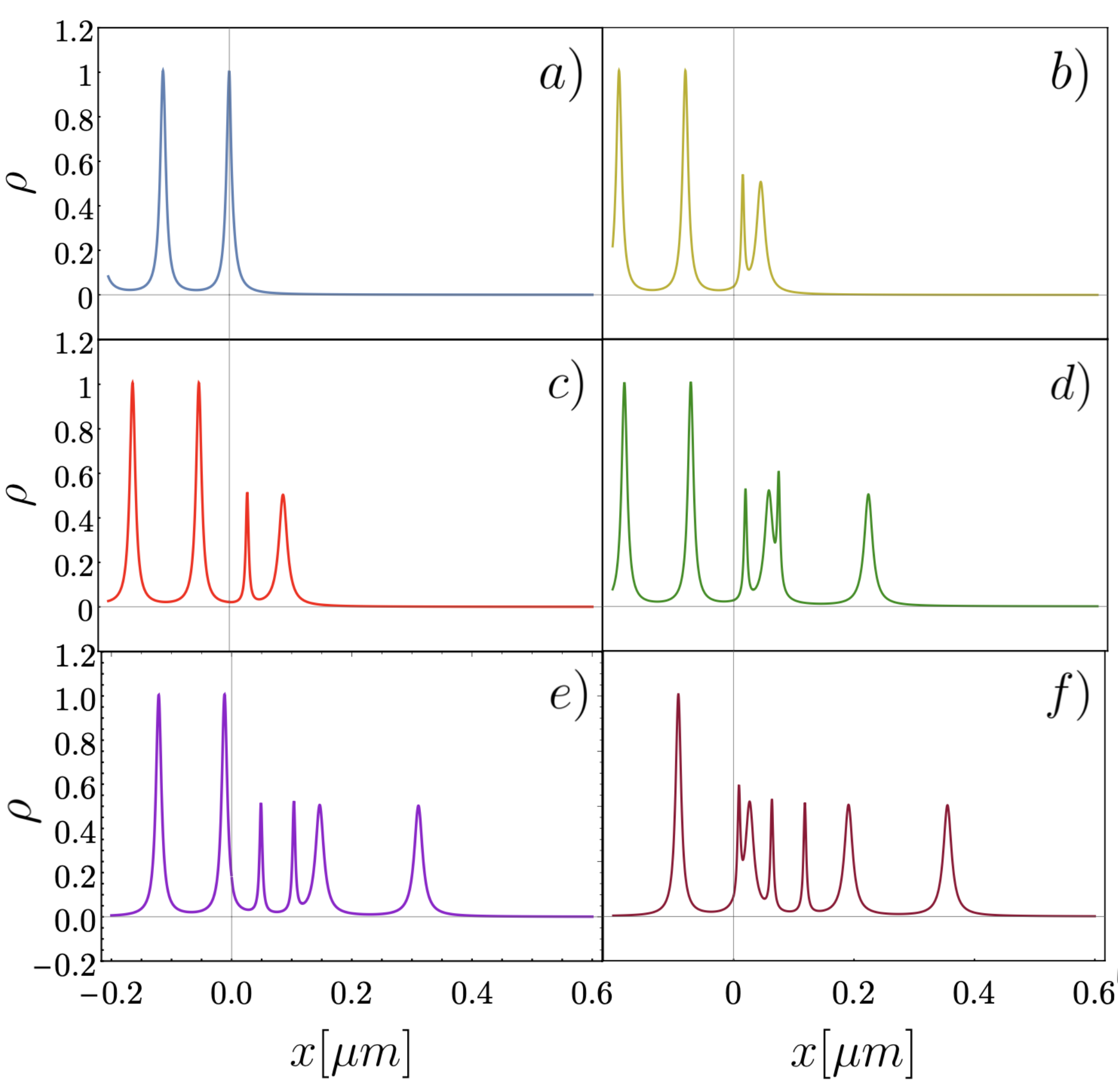}
\caption{\label{fig:fig8} Visualization of single electron charge density propagation along an edge channel. In a) a Leviton enter the interacting region $x>0$, in b) it is clear how the Coulomb coupling mechanism works: the fast mode and the slow one start separating. In d) the green curve describes the case when a second Leviton enter the interacting region: its fast mode reaches the slow mode of the previous period, recreating a purely electronic wave-packet. By repeating this mechanism for all the periods one can justify Eq.~\eqref{eq:long} and the relations between the lengths ratio and the periodic windows.}
\end{figure}
These two conditions lead to the same result, therefore we can focus on the first one
\begin{equation}
    \sum_k \tan^{-1}\!\!\left[\frac{t-k \mathcal{T}-\tau_\rho^A}{\tau_0} \right]\!=\!\sum_{k'} \tan^{-1}\!\!\left[\frac{t-k' \mathcal{T}-\tau_\rho^B+\delta}{\tau_0} \right]
\end{equation}
which yields
\begin{equation}
    (k'-k)\mathcal{T}=\tau^{A}_\rho-\tau^{B}_\rho+\delta.
    \label{eq:qk}
\end{equation}
Because we want to study how the minimum of central dip varies, the delay $\delta$ has to be fixed by
\begin{equation}
\delta_{\mathrm{cd}}=\frac{\tau^{B}_\rho+\tau^{B}_\sigma-\tau^{A}_\rho-\tau^{A}_\sigma}{2}.
\end{equation}
By substituting this expression into Eq.~\eqref{eq:qk} and recalling that $\tau^{B}_{\rho/\sigma}=\tau^{A}_{\rho/\sigma}\,L_{B}/L_{A}$, we arrive at the result in Eq.~\eqref{eq:long}.

An interesting additional feature for $\mathcal{R}(\delta_{\mathrm{cd}})$ is obtained when $\theta=\pi/4$. Due to the fact that in this case $\sin\theta=\cos\theta$, there is another possibility to fulfill $\varphi_A(t)=\varphi_B(t+\delta)$, namely 
\begin{equation}
    \varphi(t-\tau^{A}_{\rho/\sigma})=\varphi(t-\tau^{B}_{\sigma/\rho}+\delta).
\end{equation}
Again, this two conditions lead to the same result, which reads
\begin{equation}
    \frac{L_B}{L_A}=\frac{2(k'-k)\mathcal{T}}{\tau^{A}_\sigma-\tau^{A}_\rho}-1.
    \label{eq:zero-2}
\end{equation}
This analysis clearly shows that the zeros in $\mathcal{R}(\delta_{\mathrm{cd}})$ described by Eq.~\eqref{eq:long} are stable with respect to the change of the interaction strength, while those described by Eq.~\eqref{eq:zero-2} are only present at maximal coupling (see also Fig.~\ref{fig:fig7}).
In order to further characterize the physics behind this phenomenology we have reported in Fig.~\ref{fig:fig8} some snapshots of the evolution of the particle density, showing how Levitons emitted in different periods of the drive can recombine due to the interaction-induced fractionalization process.


\nocite{*}


\begin{thebibliography}{}
%
\bibitem{Grenier11} C. Grenier, R. Herv\'e, G. F\`eve, and P. Degiovanni, Mod. Phys. Lett. B \textbf{25}, 1053 (2011).
%
\bibitem{Bocquillon14} E. Bocquillon, V. Freulon, F. D. Parmentier, J.-M. Berroir, B. Pla\c cais, C.Wahl, J. Rech, T. Jonckheere, T. Martin, C. Grenier,
D. Ferraro, P. Degiovanni, and G. F\`eve, Ann. Phys. (Berlin) \textbf{526}, 1 (2014).
%
\bibitem{Bauerle18} C. B\"auerle, D. C. Glattli, T. Meunier, F. Portier, P. Roche, P. Roulleau, S. Takada, and X. Waintal, Rep. Prog. Phys. \textbf{81}, 056503 (2018).
%
\bibitem{Hanbury56} R. Hanbury Brown and R. Q. Twiss, Nature (London) \textbf{177}, 27 (1956).
%
\bibitem{Hong87} C. K. Hong, Z. Y. Ou, and L. Mandel,
Phys. Rev. Lett. \textbf{59}, 2044 (1987).
%
\bibitem{Bocquillon12} E. Bocquillon, F. D. Parmentier, C. Grenier, J.-M. Berroir, P. Degiovanni, D. C. Glattli, B. Pla\c cais, A. Cavanna, Y. Jin, and G. F\`eve, Phys. Rev. Lett. \textbf{108}, 196803 (2012).
%
\bibitem{Bocquillon13b} E. Bocquillon, V. Freulon, J.-M. Berroir, P. Degiovanni, B.Pla\c cais, A. Cavanna, Y. Jin, and G. F\`eve, Science \textbf{339}, 1054 (2013).
%
\bibitem{Feve07} G. F\`eve, A. Mah\'e, J.-M. Berroir, T. Kontos, B. Pla\c cais, D. C. Glattli, A. Cavanna, B. Etienne, Y. Jin, Science \textbf{316}, 5828 (2007).
%
\bibitem{Mahe10} A. Mah\'e, F. D. Parmentier, E. Bocquillon, J.-M. Berroir, D. C. Glattli, T. Kontos, B. Pla\c cais, G. F\`eve, A. Cavanna, and Y. Jin,
Phys. Rev. B \textbf{82}, 201309(R) (2010).
%
\bibitem{Parmentier12} F. D. Parmentier, E. Bocquillon, J.-M. Berroir, D. C. Glattli, B. Pla\c cais, G. F\`eve, M. Albert, C. Flindt, and M. B\"uttiker, Phys. Rev. B \textbf{85}, 165438 (2012).
%
\bibitem{Buttiker93} M. B\"uttiker, H. Thomas, and A. Pr\^etre, Phys. Lett. A \textbf{180}, 364 (1993).
%
\bibitem{Ol'khovskaya08} S. Ol'khovskaya, J. Splettstoesser, M. Moskalets, M. B\"uttiker, Phys. Rev. Lett. \textbf{101}, 166802 (2008).
\bibitem{Moskalets11} M. Moskalets and M. B\"uttiker, Phys. Rev. B \textbf{83}, 035316 (2011). 
%
\bibitem{Jonckheere12} T. Jonckheere, J. Rech, C. Wahl, and T. Martin, Phys. Rev. B \textbf{86}, 125425 (2012).
%
\bibitem{Haack13} G. Haack, M. Moskalets, M. B\"uttiker, Phys. Rev. B \textbf{87}, 201302(R) (2013).
%

%
\bibitem{Bocquillon13} E. Bocquillon, V. Freulon, J.-M. Berroir, P. Degiovanni, B. Pla\c cais, A. Cavanna, Y. Jin, and
G. F\`eve, Nat. Comms. \textbf{4}, 1839 (2013).
%
\bibitem{Moskalets13} M. Moskalets, G. Haack, and M. B\"uttiker, Phys. Rev. B
\textbf{87}, 125429 (2013).
%
\bibitem{Dashti19} N. Dashti, M. Misiorny, S. Kheradsoud, P. Samuelsson, and J. Splettstoesser, Phys. Rev. B \textbf{100}, 035405 (2019).
%

\bibitem{Wahl14} C. Wahl, J. Rech, T. Jonckheere, and T. Martin, Phys. Rev. Lett. \textbf{112}, 046802 (2014).
%
\bibitem{Marguerite16} A. Marguerite, C. Cabart, C. Wahl, B. Roussel, V. Freulon, D. Ferraro, Ch. Grenier, J.-M. Berroir, B. Pla\c cais, T. Jonckheere, J. Rech, T. Martin, P. Degiovanni, A. Cavanna, Y. Jin, and G. F\`eve, Phys. Rev. B \textbf{94}, 115311 (2016). 
%
\bibitem{Freulon15} V. Freulon, A. Marguerite, J.-M. Berroir, B. Pla\c cais, A. Cavanna,
Y. Jin, and G. F\`eve, Nat. Commun. \textbf{6}, 6854 (2015).
%
\bibitem{Misiorny18} M. Misiorny, G. F\`eve, J. Splettstoesser,  Phys. Rev. B \textbf{97}, 075426 (2018).
%
\bibitem{Levitov96} L. S. Levitov, H. Lee, and G. B. Lesovik, J. Math. Phys. \textbf{37}, 4845 (1996).
%
\bibitem{Ivanov97} D. A. Ivanov, H. W. Lee, and L. S. Levitov, Phys. Rev. B 56, 6839 (1997).
%
\bibitem{Keeling06} J. Keeling, I. Klich, and L. S. Levitov, Phys. Rev. Lett. \textbf{97}, 116403 (2006).
%
\bibitem{Dubois13} J. Dubois, T. Jullien, F. Portier, P. Roche, A. Cavanna, Y. Jin, W. Wegscheider, P. Roulleau, and D. C. Glattli, Nature (London) \textbf{502}, 659 (2013).
%
\bibitem{Glattli16} D. C. Glattli and P. Roulleau, Phys. E (Amsterdam, Neth.) \textbf{76}, 216 (2016).
%
\bibitem{Ferraro14} D. Ferraro, B. Roussel, C. Cabart, E. Thibierge, G. F\`eve, C. Grenier and P. Degiovanni, Phys. Rev. Lett. \textbf{113}, 166403 (2014). 
%
\bibitem{Acciai19} M. Acciai, F. Ronetti, D. Ferraro, J. Rech, T. Jonckheere, M. Sassetti, T. Martin, Phys. Rev. B \textbf{100}, 085418 (2019).
%
\bibitem{Ronetti20} F. Ronetti, M. Carrega, M. Sassetti, Phys. Rev. Research \textbf{2}, 013203 (2020).
%
\bibitem{Safi10} I. Safi and E. Sukhorukov, Europhys. Lett. \textbf{91}, 67008 (2010).
%
\bibitem{Rech17} J. Rech, D. Ferraro, T. Jonckheere, L. Vannucci, M. Sassetti, and T. Martin, Phys. Rev. Lett. \textbf{118}, 076801 (2017).
%
\bibitem{Vannucci17} L. Vannucci, F. Ronetti, J. Rech, D Ferraro, T. Jonckheere, T. Martin, and M. Sassetti, Phys. Rev. B \textbf{95}, 245415 (2017).
%
\bibitem{Ronetti18} F. Ronetti, L. Vannucci, D. Ferraro, T. Jonckheere, J. Rech, T. Martin, and M. Sassetti, Phys. Rev. B \textbf{98}, 075401 (2018).
%
\bibitem{Ferraro18b} D. Ferraro, F. Ronetti, L. Vannucci, M. Acciai, J. Rech, T. Jockheere, T. Martin, and M. Sassetti,
Eur. Phys. J. Special Topics \textbf{227}, 1345 (2018).
%
\bibitem{Ronetti19} F. Ronetti, L. Vannucci, D. Ferraro, T. Jonckheere, J. Rech, T. Martin, and M. Sassetti, Phys. Rev. B \textbf{99}, 205406 (2019).
%
\bibitem{Cabart18} C. Cabart,  B. Roussel, G. F\`eve, C. Grenier and P. Degiovanni, Phys. Rev. B \textbf{98}, 155302 (2018). 
%
\bibitem{Acciai18} M. Acciai, M. Carrega, J. Rech, T. Jonckheere, T. Martin, and M. Sassetti, Phys. Rev. B \textbf{98}, 035426 (2018).
%
\bibitem{Grenier13} C. Grenier, J. Dubois, T. Jullien, P. Roulleau, D. C. Glattli, and P. Degiovanni, Phys. Rev. B \textbf{88}, 085302 (2013).
%
\bibitem{Sukhorukov07} E. V. Sukhorukov and V. V. Cheianov, Phys. Rev. Lett. \textbf{99}, 156801 (2007).
%
\bibitem{Levkivskyi08} I. P. Levkivskyi and E. V. Sukhorukov, Phys. Rev. B \textbf{78}, 045322 (2008).
%
\bibitem{Degiovanni10} P. Degiovanni, C. Grenier, G. F\`eve, C. Altimiras, H. le Sueur, and F. Pierre, Phys. Rev. B \textbf{81}, 121302 (2010).
%
\bibitem{Ferraro17} D. Ferraro and E. V. Sukhorukov, SciPost. Phys. \textbf{3}, 014 (2017).
%
\bibitem{Hashisaka17}  M. Hashisaka, N. Hiyama, T. Akiho, K. Muraki, and T. Fujisawa,
Nat. Phys. \textbf{13}, 559 (2017).
%
\bibitem{Hashisaka18} M. Hashisaka and T. Fujisawa, 
Reviews in Physics \textbf{3}, 32 (2018)
%
\bibitem{Wen95} X.-G. Wen, Adv. Phys. \textbf{44}, 405 (1995).
%
\bibitem{Miranda03} E. Miranda, Braz. J. Phys. \textbf{33}, 3 (2003).
%
\bibitem{Braggio12} A. Braggio, D. Ferraro, M. Carrega, N. Magnoli, and M. Sassetti, New J. Phys. \textbf{14}, 093032 (2012).
%
\bibitem{Kovrizhin12} D. L. Kovrizhin, J. T. Chalker, Phys. Rev. Lett. \textbf{109}, 106403 (2012).
%
\bibitem{Inoue14} H. Inoue, A. Grivnin, N. Ofek, I. Neder, M. Heiblum, V. Umansky, and D. Mahalu, Phys. Rev. Lett. \textbf{112}, 166801 (2014).
%
\bibitem{Rodriguez20} R. H. Rodriguez, F. D. Parmentier, D. Ferraro, P. Roulleau, U. Gennser, A. Cavanna, M. Sassetti, F. Portier, D. Mailly, and P. Roche, Nature Comm. \textbf{11}, 2426 (2020). 
%
\bibitem{Sukhorukov15} E. V. Sukhorukov, Physica E \textbf{77}, 191 (2016).
%
\bibitem{Dubois13b} J. Dubois, T. Jullien, C. Grenier, P. Degiovanni, P. Roulleau, and D. C. Glattli, Phys. Rev. B \textbf{88}, 085301 (2013).
%
\bibitem{Safi99} I. Safi, Eur. Phys. J. D \textbf{12}, 451 (1999).
%

\bibitem{Altimiras10} C. Altimiras, H. le Sueur, U. Gennser, A. Cavanna, D. Mailly, and F. Pierre, Nature Phys, \textbf{6}, 34 (2010).
%
\bibitem{leSueur10} H. le Sueur, C. Altimiras, U. Gennser, A. Cavanna, D. Mailly, and F. Pierre, Phys. Rev. Lett. \textbf{105}, 056803 (2010).

%
\bibitem{Blanter00} Y. M. Blanter and M. B\"uttiker, Phys. Rep. \textbf{336}, 1 (2000).
%
\bibitem{Martin05} T. Martin, \emph{Les Houches Session LXXXI} edited by H. Bouchiat S. Gu\'eron, Y. Gefen, G. Montambaux, and J. Dalibard (Elsevier, Amsterdam, 2005).
%
\bibitem{Ferraro14b} D. Ferraro, M. Carrega, A. Braggio, and M. Sassetti, New J. Phys. \textbf{16}, 043018 (2014).
%
\bibitem{Ferraro13} D. Ferraro, A. Feller, A. Ghibaudo, E. Thibierge, E. Bocquillon, G. F\`eve, C. Grenier, and P. Degiovanni, Phys. Rev. B \textbf{88}, 205303 (2013).
%
\bibitem{Grenier112} C. Grenier, R. Herv\'e, E. Bocquillon, F. D. Parmentier, B. Pla\c cais, J.-M. Berroir, G. F\`eve, P. Degiovanni, New J. Phys. \textbf{13}, 093007 (2011).
%

\bibitem{Moskalets16} M. Moskalets, G. Haack, Physica E \textbf{75}, 358 (2016).
%
\bibitem{Ferraro18} D. Ferraro, F. Ronetti, J. Rech, T. Jonckheere, M. Sassetti, and T. Martin, Phys. Rev. B \textbf{97}, 155135 (2018).

\end{thebibliography}
\end{document}